%% file: cpvmix-proc.tex
\documentclass[12pt]{article}
\usepackage{graphicx}

\newcommand\pubnumber{}
\newcommand\pubdate{\today}

\def\cbpf{Centro Brasileiro de Pesquisas F\'{\i}sicas \\
Rio de Janeiro,   Brazil}
\def\support{\footnote{on behalf of the LHCb collaboration.}}

\def\Title#1{\begin{center} {\Large #1 } \end{center}}
\def\Author#1{\begin{center}{ \sc #1} \end{center}}
\def\Address#1{\begin{center}{ \it #1} \end{center}}

\newcommand\pubblock{\rightline{\begin{tabular}{l} \pubnumber\\
         \pubdate  \end{tabular}}}
\newenvironment{Abstract}{\begin{quotation}  }{\end{quotation}}
\newenvironment{Presented}{\begin{quotation} \begin{center} 
             PRESENTED AT\end{center}\bigskip 
      \begin{center}\begin{large}}{\end{large}\end{center} \end{quotation}}
\def\Acknowledgements{\bigskip  \bigskip \begin{center} \begin{large}
             \bf ACKNOWLEDGEMENTS \end{large}\end{center}}

\input econfmacros.tex

\begin{document}
\begin{titlepage}
\pubblock

\vfill
\Title{Charm mixing and CP violation}
\vfill
\Author{ Alberto C. dos Reis\support}
\Address{\cbpf}
\vfill
\begin{Abstract}
Experimental results on charm mixing and CP violation searches  are reviewed.
This paper focus on results released after FPCP 2013.
\end{Abstract}
\vfill
\begin{Presented}
Flavor Physics and CP Violation (FPCP-2014)\\
Marseille, France, May 26-30, 2014

\end{Presented}
\vfill
\end{titlepage}
\def\thefootnote{\fnsymbol{footnote}}
\setcounter{footnote}{0}

\section{Introduction}

In 2014 we celebrate the 40th anniversary of the discovery of the charm quark 
--- the "November revolution". It took over 30 years of experimental efforts 
to achieve the necessary sensitivity for the establishment of charm 
mixing\footnote{In the literature the terms "mixing" and "oscillations" 
are used interchangeably.
Mixing arises from the fact that mass and flavour eigenstates are not identical. 
One manifestation of mixing is the existence of mass eigenstates with different
lifetimes. Mixing also induces flavour oscillations, in-flight transitions between 
a neutral meson and its antiparticle. Oscillations are characterized by a 
sinusoidal behavior of the time evolution of neutral system.
In this review the notation "charm mixing" is used as a generic designation 
of both mixing and oscillations.}, which happened in 2008 when of results 
from different experiments~\cite{belle0mix,babar0mix,cdf.mix,hfag} were 
combined.

This year we also celebrate the 50th anniversary of the discovery 
of \cp violation. While the \cp violation is well established in decays of
neutral kaons and $B$ mesons, in charm it has not been observed yet.

Mixing and \cp violation in charm are regarded as promising fields
in the search for new physics, especially after the commissioning of the LHC.
The interpretation of the experimental results, however, is still problematic. Predictions of the Standard Model (SM) contribution to the mixing rate and
\cp\ asymmetries suffer from uncertainties on the hadronic matrix elements.

In this paper we review the main experimental results on charm mixing and 
\cp violation searches that were released after the 2013 edition of FPCP, 
in B\'uzios.

\section{Time-dependent measurements}

\subsection{Mixing and CP violation search with WS \ $D^0\!\to\!K\pi$}

\begin{figure}[htb]
\centering
\includegraphics[width=9cm]{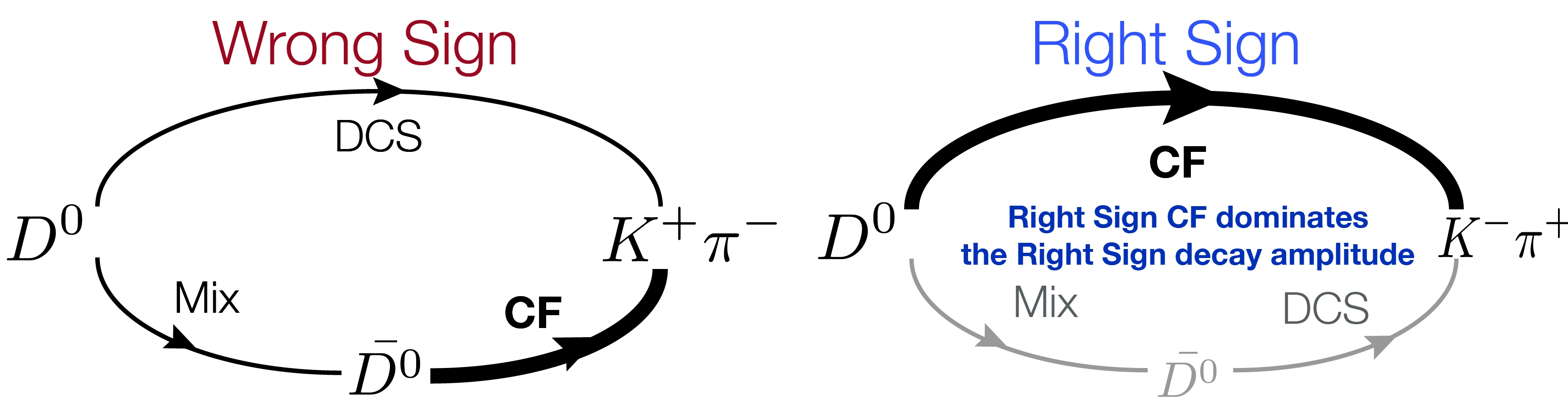}
\caption{The two paths for WS $D^0 \to K^+\pi^-$ decays. The RS 
$D^0 \to K^-\pi^+$ decay is completely dominated by the CF transition.}
\label{fig:wsrs}
\end{figure}

The measurement of the time-dependent ratio of the "wrong-sign" $D^0 \to K^+\pi^-$
(WS) to the "right-sign" (RS) $D^0 \to K^-\pi^+$ decay rates is the most 
sensitive method for observing \ddbar oscillations. The RS decay rate is, 
for any practical purpose, given by the Cabibbo favoured
(CF) transition $D^0 \to K^-\pi^+$, as illustrated in Fig. \ref{fig:wsrs}. 
There is a negligibly small contribution from processes where a net oscillation to
a $\overline{D}^0$ is followed by the doubly Cabibbo suppressed
(DCS) transition $\overline{D}^0 \to K^-\pi^+$. 
For the WS decay rate, on the other hand, there are two competing contributions: 
the direct DCS transition $D^0 \to K^+\pi^-$; a net oscillation to a
$\overline{D}^0$ followed by the CF transition $\overline{D}^0 \to K^+\pi^-$. 
The concurrence of these two paths allows one to extract the mixing parameters.  

An extra bonus of this method comes from the fact that the WS and RS decays  
are expected to have the same decay-time acceptance, so the ratio between
WS and RS events cancels most of the systematic uncertainties affecting the determination of the yields as a function of the decay time.

The mixing rate is characterized by two dimensionless parameters $x$ and $y$,
\begin{equation}
x\equiv \frac{m_1-m_2}{\Gamma_D}=\frac{\Delta m}{\Gamma_D}, \hskip 1cm 
y\equiv \frac{\Gamma_1-\Gamma_2}{2\Gamma_D}=\frac{\Delta \Gamma}{2\Gamma_D},
\end{equation}
where \ $m_{1,2}$ \ and \ $\Gamma_{1,2}$ are the masses and widths of the mass
eigenstates,  and \ $\Gamma_D\equiv  (\Gamma_1+\Gamma_2)/2$ \  is the inverse 
of the average $D^0$ lifetime.

In the $D^0\to K\pi$ method one has to account for the relative strong phase 
between the amplitudes for the CF and DCS decays,
\[
\frac{A_{K^+\pi^-}}{\overline{A}_{K^+\pi^-}} = -\sqrt{R_D} \ e^{-i\delta_{K\pi}}, 
\]

The mixing parameters extracted with this method are rotated with respect
to those defined by eq. 1,
\[
x'= x\cos\delta_{K\pi} + y \sin \delta_{K\pi},\hskip .8cm
y'= y\cos\delta_{K\pi} - x \sin \delta_{K\pi}.
\]

The relative phase $\delta_{K\pi}$ is an external input that is measured 
in $e^+e^-$ storage rings \cite{cleo-cor,bes-cor} where,
at a center of mass energy corresponding to the $\psi(3770)$ resonance, a $D\!\overline{D}$ pair is produced in a quantum correlated state with $C=1$.  

In the limit of small mixing ($x,y\ll1$) and neglecting \cp violation, 
the time-dependent ratio of WS to RS decay rates is given by,
\begin{equation}
R(t)\approx R_D + \sqrt{R_D}y'\frac{t}{\tau} +
\frac{x'^2+y'^2}{4}\left(\frac{t}{\tau}\right)^2 ,
\label{eq:rt}
\end{equation}
where $t/\tau$ is the decay time expressed in units of the average $D^0$ 
lifetime $\tau$. The decay time $t$ is computed in the $D^0$ rest frame using 
the measured $D^0$ mass, the distance from the production to the decay 
positions and the $D^0$  momentum, \ $t=m_D\Delta \vec{X}\cdot\vec{p}/|\vec{p}|^2$. The typical decay time resolution in LHCb is approximately $0.1\tau$.

The first term in the right side of Eq. (\ref{eq:rt}) is the time-integrated 
ratio between the DCS and CF decay rates. The linear term in $t/\tau$ 
corresponds to the interference between WS decays with and without oscillations,
whereas the quadratic term accounts for the pure mixing contribution.

If \cp violation is not neglected, the time-dependent ratio between WS and RS 
decays may differ for $D^0$ and $\overline{D}^0$. The WS-to-RS ratios for 
initially produced $D^0$ and $\overline{D}^0$, denoted by $R^+(t)$ and $R^-(t)$,
respectively, are functions of six independent mixing parameters 
$(R_D^{\pm},x'^{2\pm},y'^{\pm})$. Different values for ($x'^{2+},y'^+$) 
and ($x'^{2-},y'^-$) imply \cp violation, either in mixing ($|q/p|\neq 1$) or
in the interference between the amplitudes for decays with and without a net
oscillation  
($\phi \equiv\mathrm{arg}[qA(\overline{D}^0\!\to\!K^-\pi^+)/
pA(D^0\!\to\!K^-\pi^+)]-\delta_{K\pi}\neq 0$).

\begin{figure}[htb]
\centering
\includegraphics[width=6.5cm]{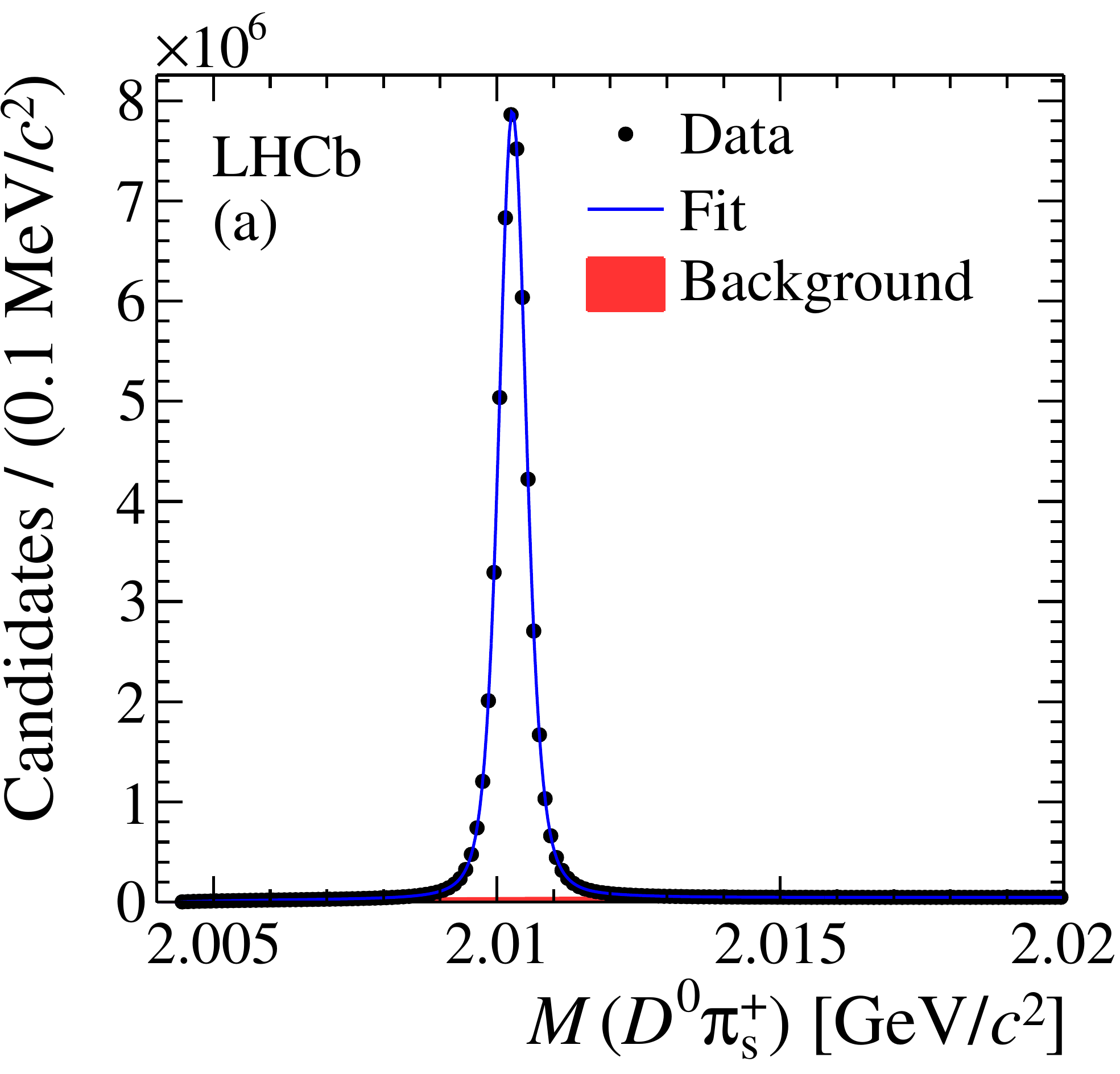}\hskip .8cm
\includegraphics[width=6.5cm]{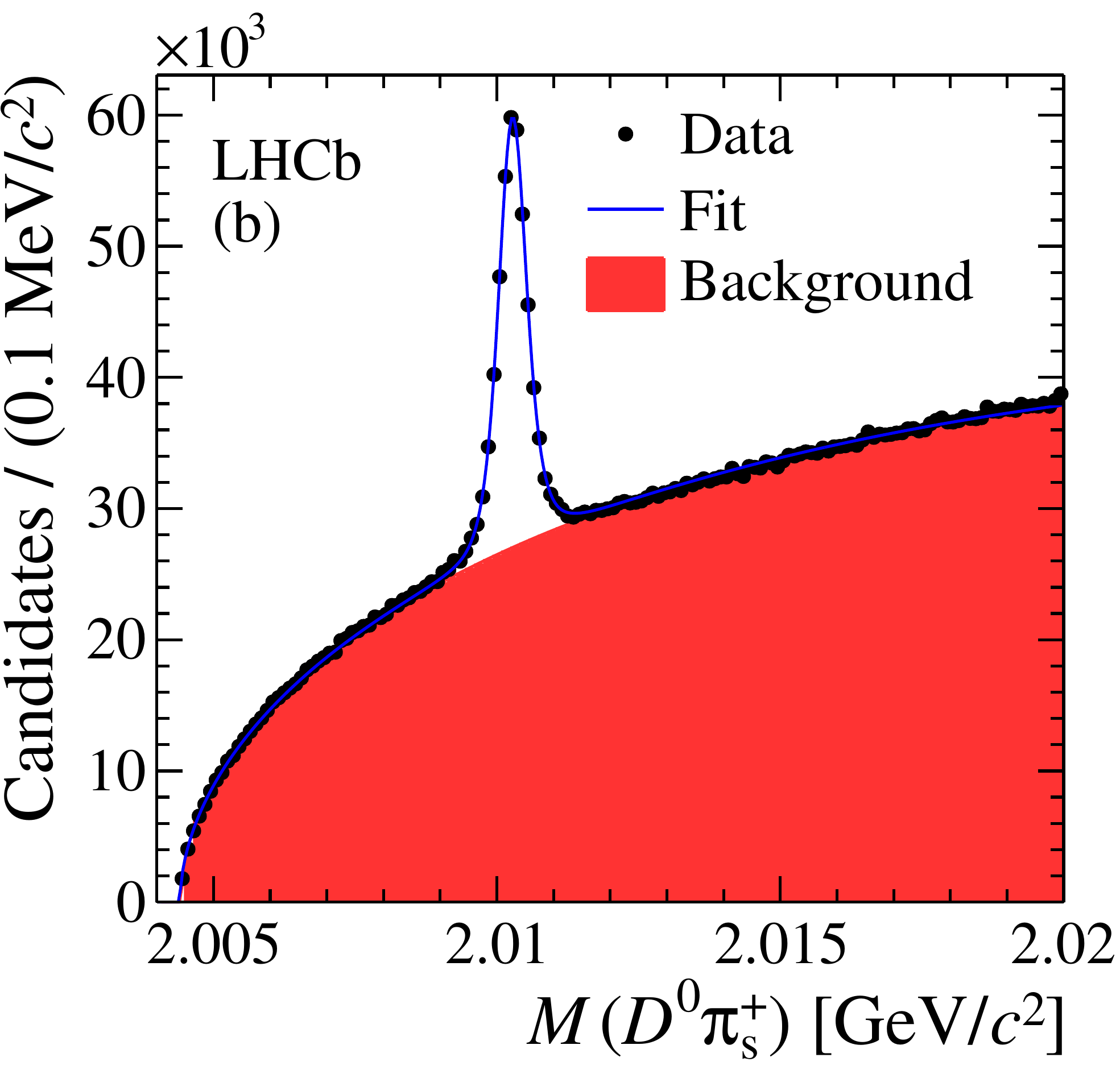}
\caption{The RS $D^0\to K^-\pi^+$ (left) and WS $\overline{D}^0\to K^+\pi^-$ 
(right) time-integrated $D^0\pi^+$ mass distributions from LHCb (3 fb$^{-1}$). 
The projection of the fits (blue line) are overlaid. In the bottom plots 
the normalized residuals between the data points and fits are shown.}
\label{fig:kpi1}
\end{figure}

Using the full data set from the 2011 and 2012 runs, LHCb reported on
an improved measurement of the {\em CP}-averaged charm mixing parameters
 and a search for  {\em CP} violation using the
$D^0\to K^+\pi^-$ decay\cite{lhcbmix2}. The measurement is based on approximately 
$2.3\times10^5$ WS  and $5.3\times10^7$ RS decays. 
The $D^0\pi^+$ invariant mass spectrum is shown in Fig.~\ref{fig:kpi1}.
The background under the WS signal is dominated by favored 
$\overline{D}^0\to K^+\pi^-$ decays associated to a random slow pion.

The data shown in  Fig.~\ref{fig:kpi1} is divided into 
thirteen bins of decay time with approximately the same number of decays. 
The RS and WS yields are determined from thirteen independent fits of the
$M(D^0\pi^+)$ distribution,  where only $D^0$ candidates with mass within
$\pm$24 MeV/$c^2$ of the nominal value are considered. 
The mixing parameters are determined by
minimizing a $\chi^2$ variable that includes\ the observed and predicted ratios, 
and terms accounting for the systematic effects
associated to candidates from $b$-hadron decays, double misidentification of
the final state particles and instrumental asymmetries in the $K^-\pi^+/K^+\pi^-$ 
reconstruction efficiency.

Fits are performed under different hypothesis: allowing for direct and
indirect \cp violation; allowing only for indirect \cp violation, which constrains $R^{\pm}_D$ to a single value; a {\em CP}-conserving fit, constraining all mixing parameters to be common to $D^0$ and $\overline{D}^0$. 

\begin{figure}[htb]
\centering
\includegraphics[width=6.5cm]{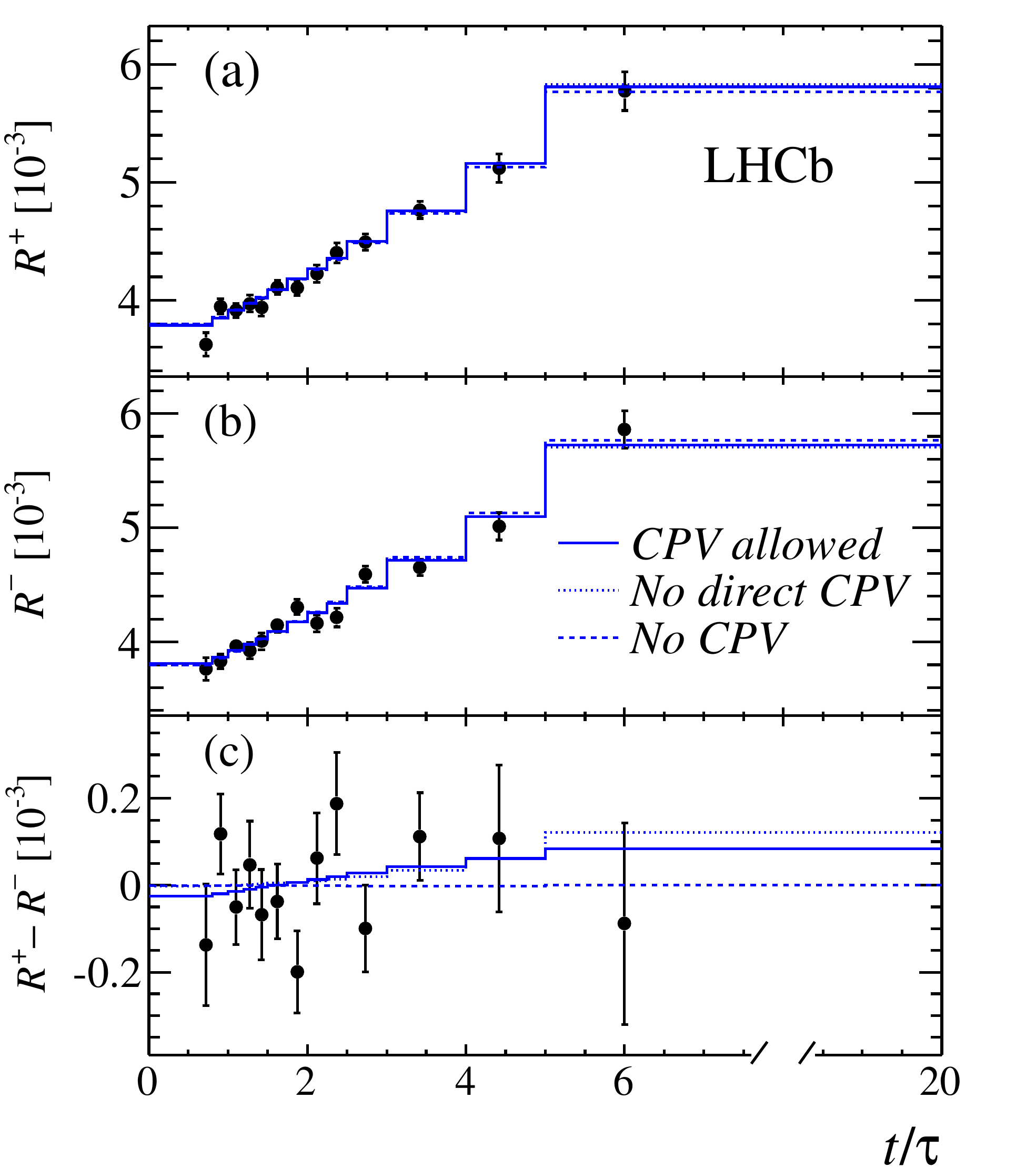}
\caption{Efficiency-corrected ratios of WS-to-RS yields, $R^+$ and $R^-$,
for $D^0$ and $\overline{D}^0$, and their difference as a function of the
decay time in units of $D^0$ lifetimes, from LHCb. Projections of the three types
of fits are superimposed.}
\label{fig:kpi2}
\end{figure}

The fit results are shown in Fig.~\ref{fig:kpi2}. The efficiency-corrected 
ratios $R^+(t)$ and $R^-(t)$ and the difference  $R^+(t)-R^-(t)$ are displayed 
(black full circles with error bars) as functions of the decay time, in units of 
$D^0$ lifetime, with the projections of the fit results (lines in blue) superimposed. The slope of the  $R^+(t)-R^-(t)$ difference is approximately
5\% of the individual slopes  of the $R^+(t)$ and $R^-(t)$ distributions and
is consistent with zero.

The signal for direct \cp violation would be a nonzero intercept at 
$t=0$ of the $R^+(t)-R^-(t)$ distribution. 
This intercept is parameterized by asymmetry $A_D$,
\[
A_D \equiv \frac{R_D^+-R_D^-}{R_D^++R_D^-}.
\]
The asymmetry  $A_D$ is measured from the fit with all forms of \cp violation 
allowed. The result is $A_D=(-0.7 \pm 1.9)\%$.

The fit results are summarized in Table \ref{tab:xykpi2} with statistical 
and systematic uncertainties, respectively.

\begin{table}[ht]
\begin{center}
\caption{Measured parameters of \ddbar mixing, obtained from the ratio of 
wrong sign to right sign $D^0\to K^+\pi^-$ decays, for different hypothesis 
on \cp symmetry.}
{\begin{tabular}{lcc} \\\hline  
Parameter  & Fit result \\
\hline
Direct and indirect \cp violation  & \\ 
                          
$R_D^+ (\times 10^{-3})$   &  3.545 $\pm$ 0.082 $\pm$ 0.048   \\ 
$y'^+  (\times 10^{-3})$   &  5.1 $\pm$ 1.2 $\pm$ 0.7         \\
$x'{2+} (\times 10^{-5})$  &  4.9 $\pm$ 6.0 $\pm$ 3.6         \\
$R_D^- (\times 10^{-3})$   &  3.591 $\pm$ 0.081 $\pm$ 0.048   \\ 
$y'^-  (\times 10^{-3})$   &  4.5 $\pm$ 1.2 $\pm$ 0.7         \\
$x'{2-} (\times 10^{-5})$  &  6.0 $\pm$ 5.8 $\pm$ 3.6         \\
$\chi^2/\mathrm{ndof}$     &   85.9/98                        \\
 & \\
Indirect \cp violation  &   \\
$R_D (\times 10^{-3})$     &  3.568 $\pm$ 0.0058 $\pm$ 0.033  \\ 
$y'^+  (\times 10^{-3})$   &  4.8 $\pm$ 0.9 $\pm$ 0.6         \\
$x'{2+} (\times 10^{-5})$  &  6.4 $\pm$ 4.7 $\pm$ 3.0         \\
$y'^-  (\times 10^{-3})$   &  4.8 $\pm$ 0.9 $\pm$ 0.6         \\
$x'{2-} (\times 10^{-5})$  &  4.6 $\pm$ 4.6 $\pm$ 3.0         \\
$\chi^2/\mathrm{ndof}$     &   86.0/99                        \\
 & \\
No \cp violation & \\
$R_D (\times 10^{-3})$    &  3.568 $\pm$ 0.0058 $\pm$ 0.033   \\ 
$y'  (\times 10^{-3})$    &  4.8 $\pm$ 0.8 $\pm$ 0.5          \\
$x'{2} (\times 10^{-5})$  &  5.5 $\pm$ 4.2 $\pm$ 2.6          \\
$\chi^2/\mathrm{ndof}$    &   86.4/101                        \\
\hline      
\end{tabular}
\protect\label{tab:xykpi2}}
\end{center}
\end{table}

\begin{figure}[htb]
\centering
\includegraphics[width=10.5cm]{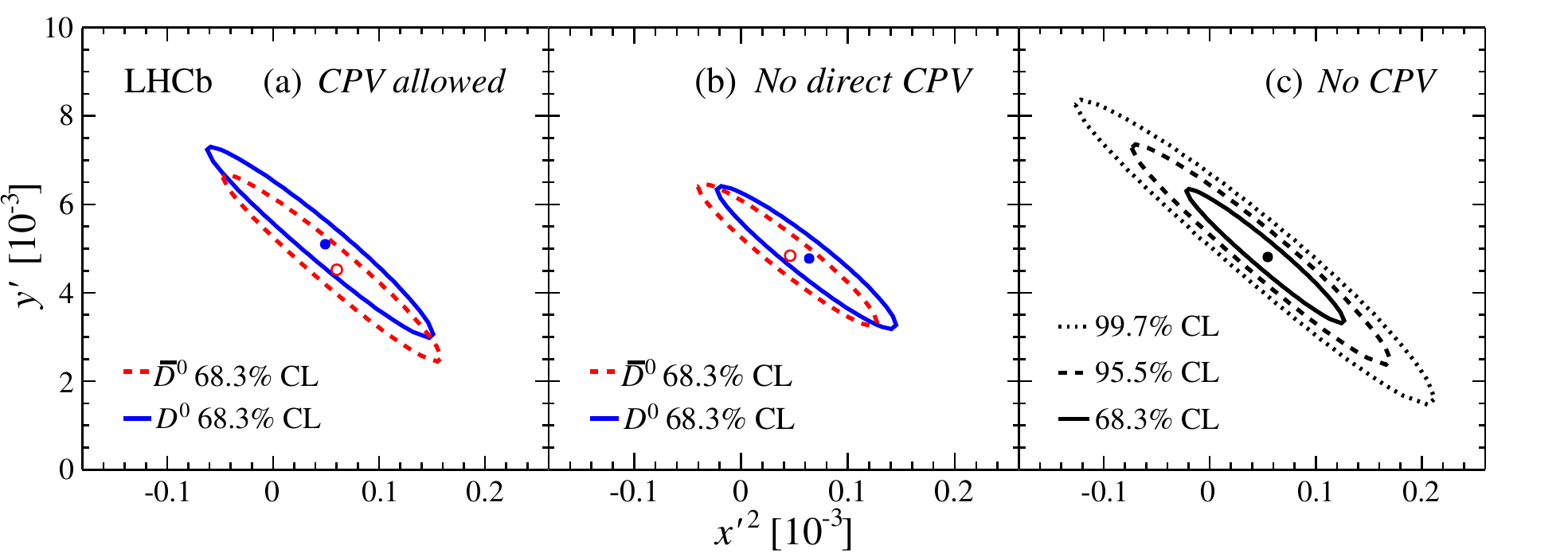}
\caption{Two-dimensional CL contours in the ($x'^2,y'$) plane, from LHCb, for the
three types of fit: (a) no restriction on \cp violation; (b) assuming no 
direct \cp violation; (c) assuming \cp conservation.}
\label{fig:kpi3}
\end{figure}

The central values and confidence regions in the $(x'^2,y')$ plane
are shown in Fig.~\ref{fig:kpi3} for the three fits. The data is compatible 
with \cp conservation.

\subsection{$A_{\Gamma}$ \ from \ $D^0 \to K^+K^-$ \ and \ $D^0 \to \pi^+\pi^-$}

Decays of $D^0$ to final states that are \cp eigenstates,
such as $D^0 \to K^-K^+$ and $D^0 \to \pi^-\pi^+$, provide an 
alternative way to search for \cpv\ in mixing. Under the assumption of 
no direct \cpv, the decay time distributions, to a good approximation,
can be described by pure exponentials with effective widths\cite{bergman},
\begin{eqnarray}
\hat{\Gamma}(D^0\!\to\!h^+h^-)&=& 
\Gamma_D \left[1+ \left|\frac{q}{p}\right|(y\cos\phi-x\sin\phi)\right], \nonumber \\
\hat{\Gamma}(\overline{D}^0\!\to\!h^+h^-)&=& 
\Gamma_D \left[1+ \left|\frac{p}{q}\right|(y\cos\phi+x\sin\phi)\right],  \nonumber\\
\hat{\Gamma}(D^0\!\to\!K^-\pi^+)&=&\hat{\Gamma}(\overline{D}^0\to K^+\pi^-) = 
\Gamma_D, 
\label{eq:agamma}
\end{eqnarray}
where $\Gamma_D$ is the average $D^0$ lifetime, $\phi$ is the relative phase between
the amplitudes for decays with and without mixing, and $h=K,\pi$.

Two observables can be built from the effective lifetimes:
\[
y_{CP} = \frac{\hat{\Gamma}_{D^0\to h^+h^-}+\hat{\Gamma}_{\overline{D}^0\to h^+h^-}}
{2\Gamma_D} - 1,
\]
and
\[
A_{\Gamma}\equiv 
\frac{\hat{\Gamma}(D^0\to h^+h^-)-\hat{\Gamma}(\overline{D}^0\to h^+h^-)}
{\hat{\Gamma}(D^0\to h^-h^+)+\hat{\Gamma}(\overline{D}^0\to h^+h^-)}
\simeq \left(\frac{1}{2}A_m y \cos\phi-x\sin\phi \right),
\]
with
\[
A_m =\frac{|q/p|^2-|p/q|^{-2}}{|q/p|^2+|p/q|^{-2}}. 
\]

Any difference between \ $\hat{\Gamma}_{D^0\to h^+h^-}$ \ and \ 
$\hat{\Gamma}_{D^0\to {K\pi}}$ \ is a signal of mixing, whereas any difference 
between $\hat{\Gamma}_{D^0\to h^+h^-}$ and $\hat{\Gamma}_{\overline{D}^0\to h^+h^-}$ \
means indirect \cpv, either through the phase $\phi$ or through $A_m$.

A measurement of $A_{\Gamma}$ using the decays $D^0 \to K^-K^+$ and $D^0 \to \pi^-\pi^+$
was performed by LHCb, with 1 fb$^{-1}$ of $pp$ collisions at 7 TeV\cite{agamma-11}. 
In this analysis flavor tagging is performed through the charge of the slow pion
in the chain $D^{*+} \to D^0\pi^+$, $D^0 \to hh$.
The reconstructed momentum of the slow pion is constrained to the $pp$ interaction
vertex, a technique that improves the  signal resolution in the mass difference 
$\Delta m = m(hh\pi)-m(hh)$ spectrum, with a consequent improvement in
the signal-to-noise ratio.

\begin{figure}[htb]
\centering
\includegraphics[width=6.5cm]{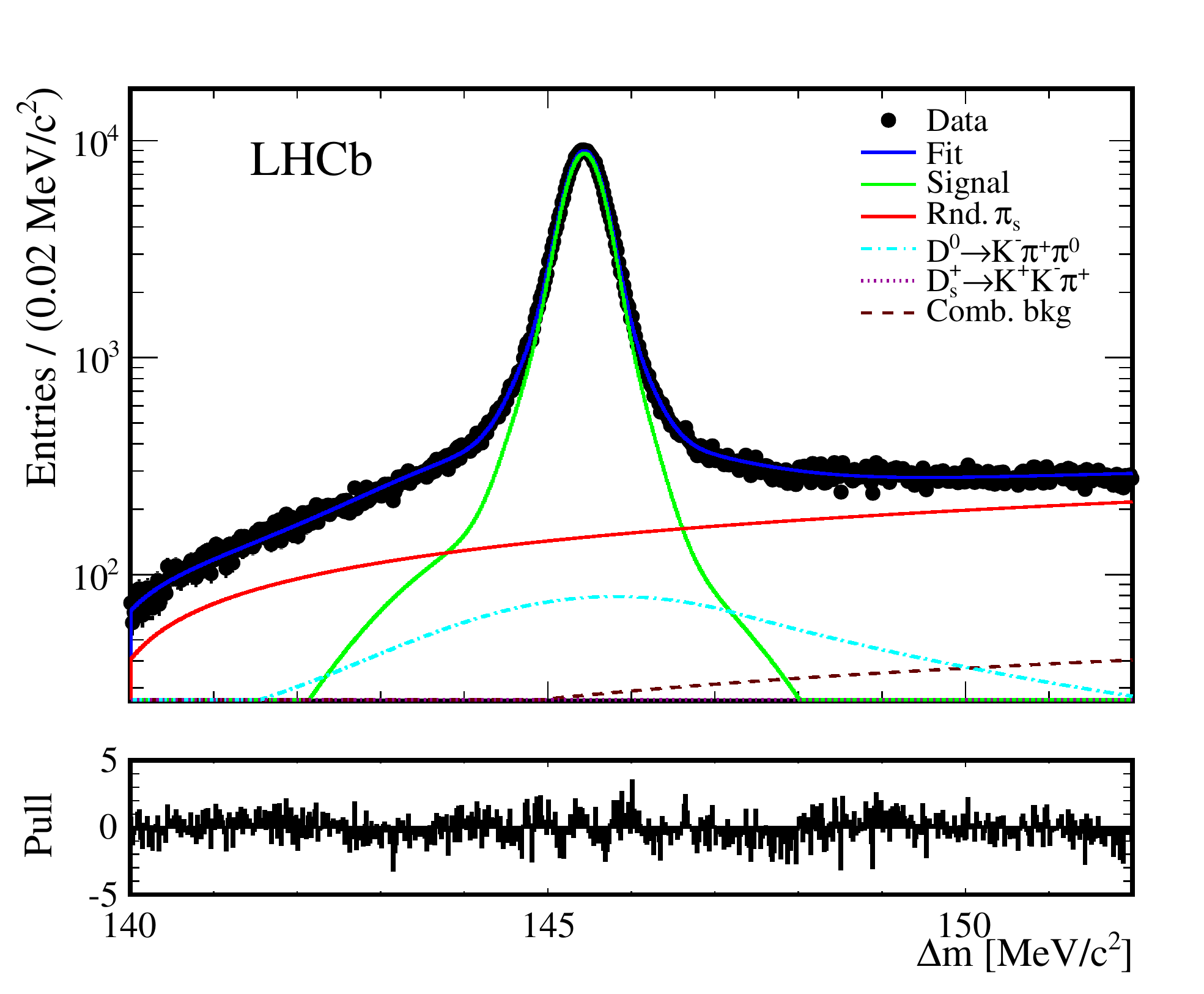}
\includegraphics[width=6.5cm]{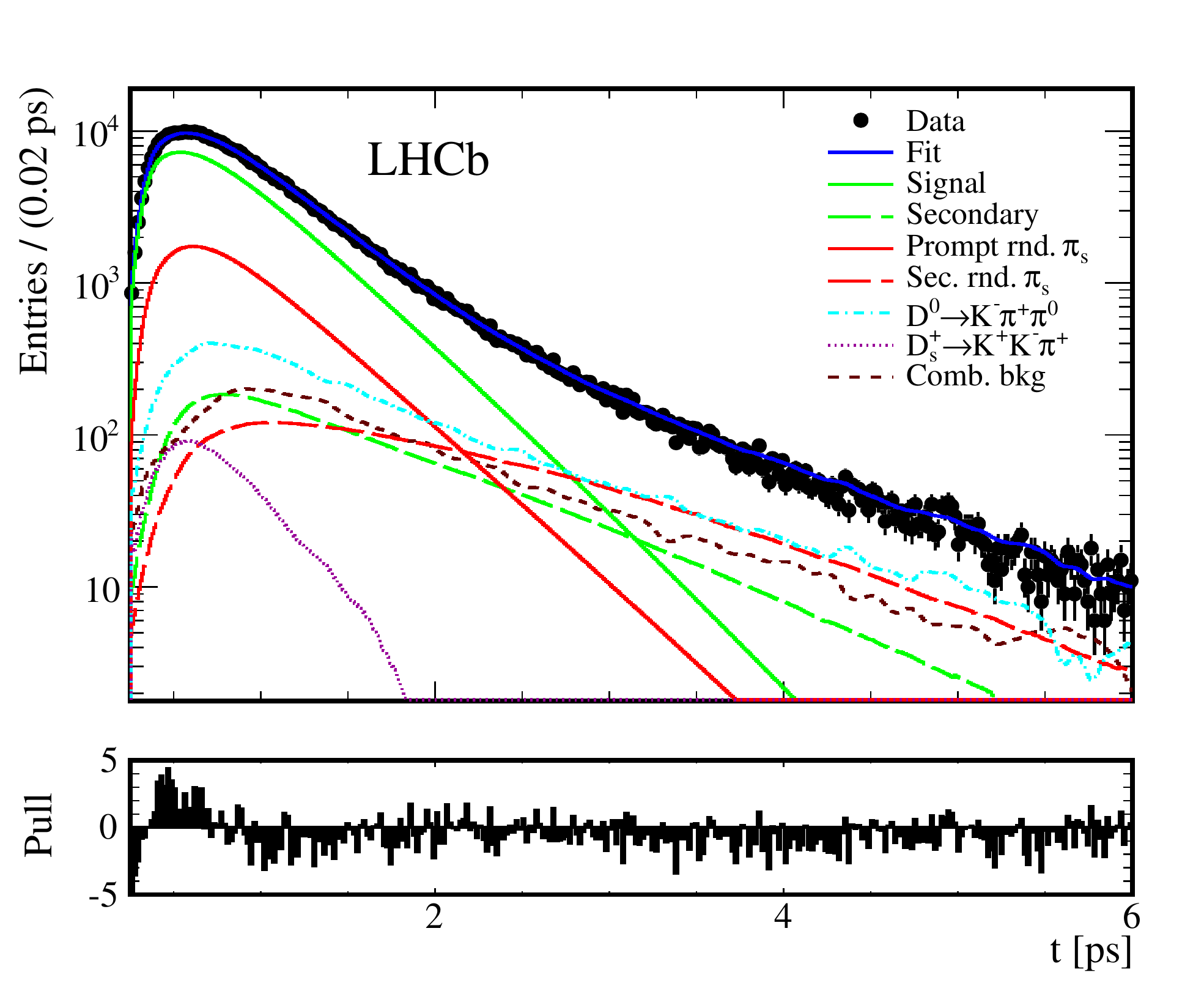}
\caption{Left: mass difference $m(K^-K^+\pi^-)-m(K^+K^-)$, from LHCb, 
with the fit result superimposed including the various background components. 
Right: the decay time distribution of $\overline{D}^0 \to K^-K^+$ candidates
with fit results superimposed.
Both plots correspond to one of the eight subsets (see text for more retails).}
\label{fig:agamma1}
\end{figure}

The selected sample contains $3.11\times 10^6$ $D^0 \to K^+K^-$ and 
$1.03\times 10^6$ $D^0 \to \pi^+\pi^-$ candidates. For each final state the
selected sample is divided into eight subsets, according to the flavor of the 
$D^0$, the magnet polarity and data taking period. The latter accounts for
important changes in the trigger configuration during the 2011 run. 
In the left plot of Fig.~\ref{fig:agamma1} 
the $\Delta m$ distribution of one of the eight subsets is shown, for the 
$D^0 \to K^-K^+$ channel.

\begin{figure}[htb]
\centering
\includegraphics[width=4.5cm]{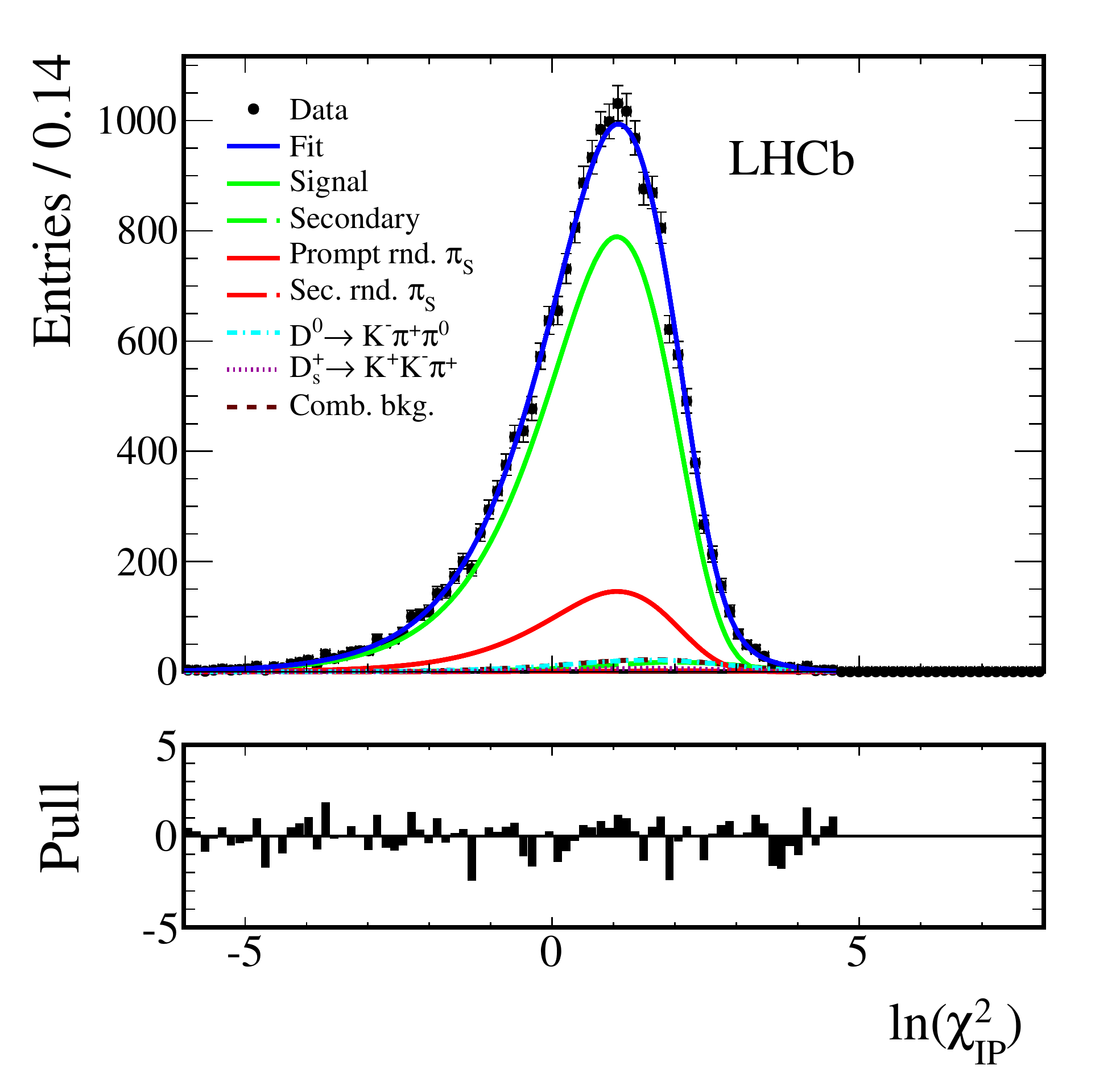}
\includegraphics[width=4.5cm]{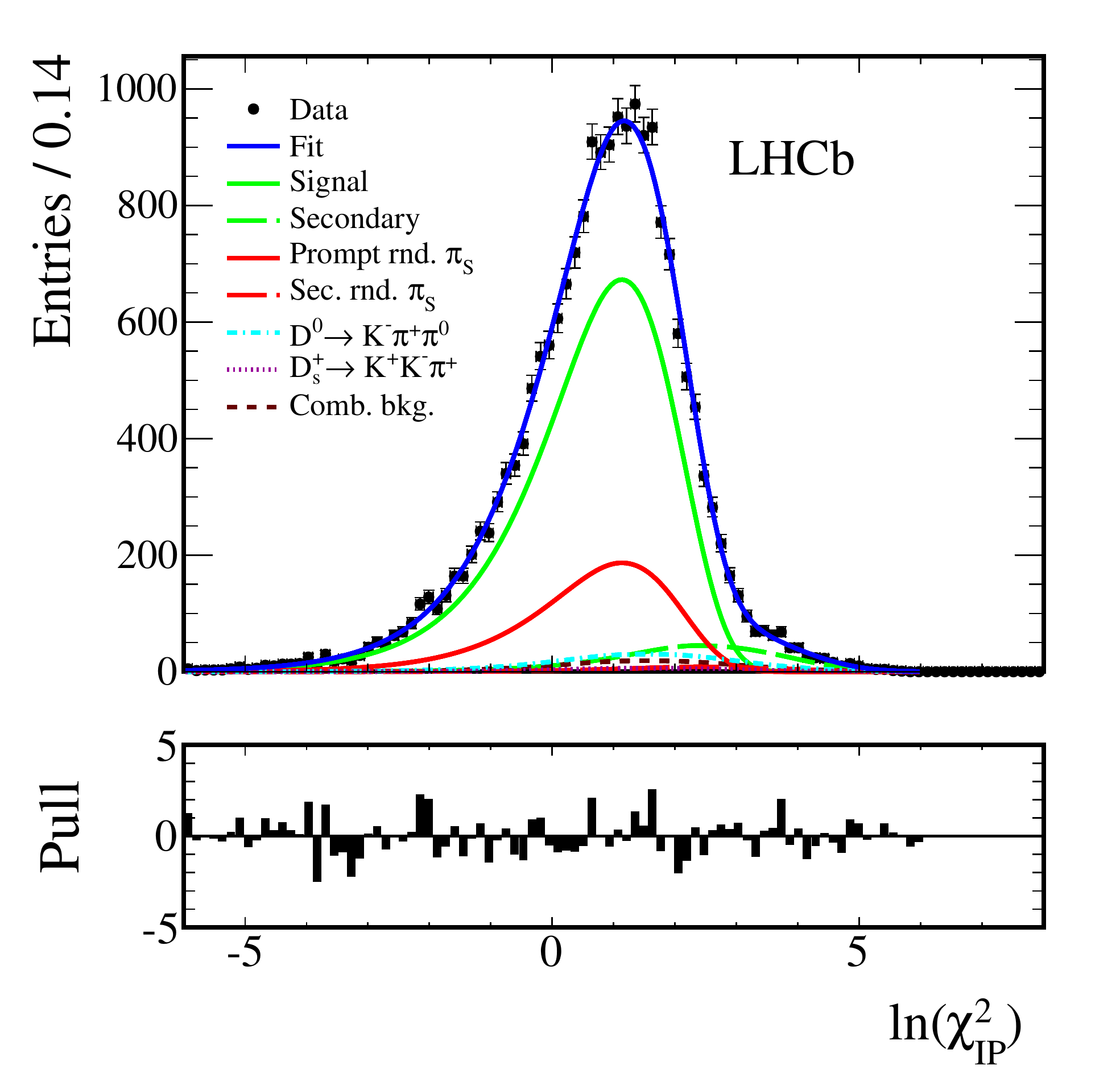}
\includegraphics[width=4.5cm]{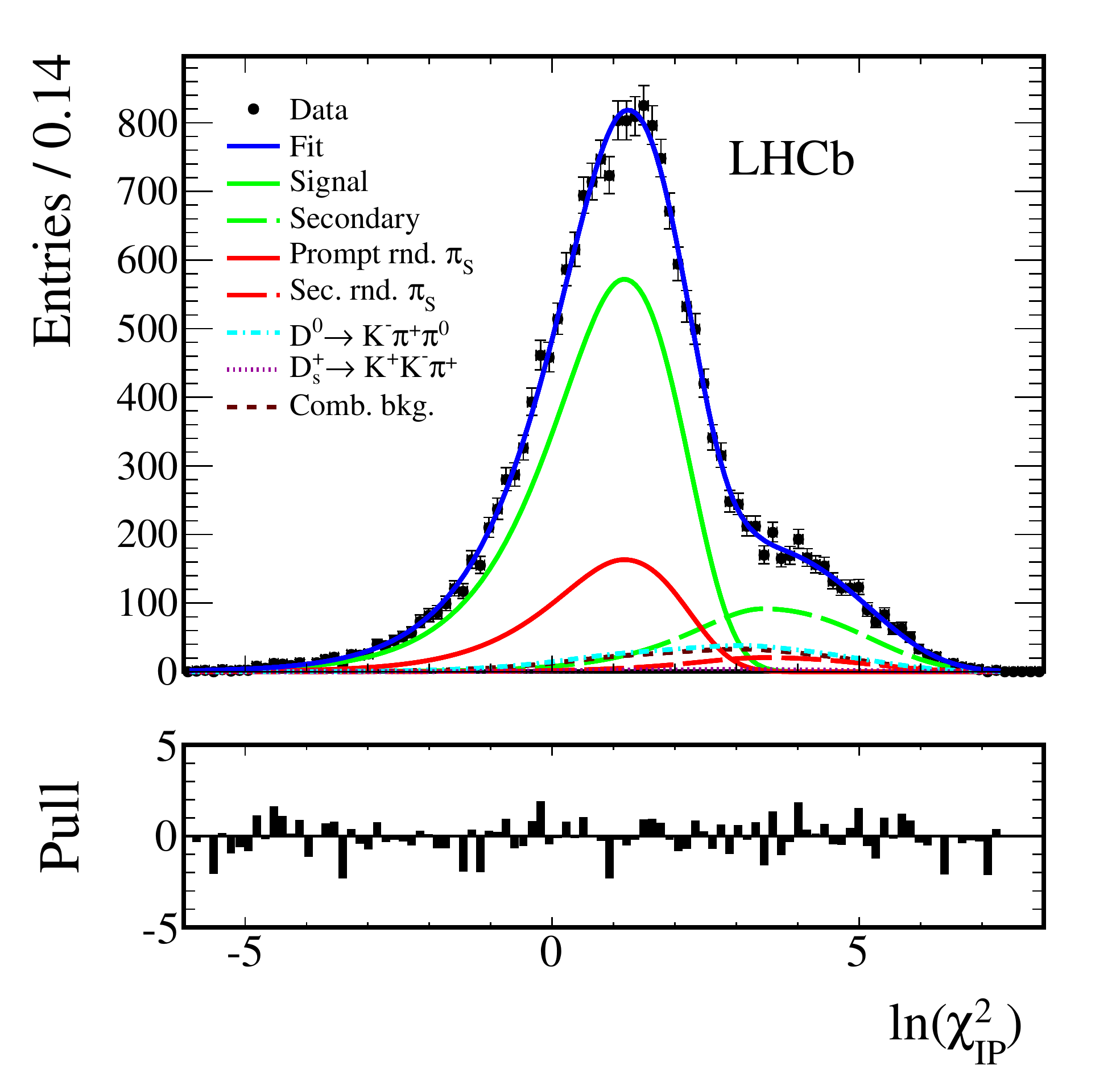}
\caption{The distribution of the logarithm of an impact parameter variable
for three $D^0$ decay time intervals. For longer decay times the contribution
from $D^0$ produced in $B$ decays becomes more important.}
\label{fig:agamma2}
\end{figure}

The determination of the effective lifetimes is performed in two stages.
In the first stage, a bi-dimensional fit to the $m(hh)$ and $\Delta m$
distributions is carried out, fixing the fractions of the signals and the main
classes of background, namely true $D^0$ associated to a random slow pion, partially
reconstructed decays and combinatorial background. 

$D^0$ candidates from decays of $b$-hadrons, or secondary $D^0$, 
is an important background that cannot 
be separated by the mass fit. Since $b$-hadron have long lifetimes, 
this background is more important at larger values of the decay time. 
The impact parameter with respect to the  primary $pp$ interaction 
vertex is used to discriminate the promptly produced $D^0$ candidates from
secondary $D^0$. In  Fig.~\ref{fig:lhcb-agamma-3} the distribution of
an impact parameter variable  is shown for three decay time intervals.

The effective lifetimes are extracted in the second stage
by a two-dimensional fit to the impact parameter and decay time distributions.
The fit result, for one of the eight subsets of the $D^0 \to K^-K^+$ sample,
is displayed in the right plot of Fig.~\ref{fig:lhcb-agamma-2a}. 

An alternative method is used as a check to the nominal unbinned fit results. The data
is divided in bins of decay time with approximately equal population. The ratio
of $\overline{D}^0$ to $D^0$, is computed in each bin and the value of 
$A_{\Gamma}$ is obtained from a linear $\chi^2$ minimization. The results obtained
with the unbinned and binned methods are in good agreement.

The results obtained by LHCb with 1.0 fb$^{-1}$ are:

\begin{eqnarray}
A_{\Gamma}(KK) =& (-0.035 \pm 0.062 \pm 0.012)\% \\
A_{\Gamma}(\pi\pi) =& (0.033 \pm 0.106 \pm 0.014)\%
\end{eqnarray}

The measured values of $A_{\Gamma}(KK)$  and $A_{\Gamma}(\pi\pi)$ are consistent
with each other and with the no \cpv hypothesis.

\subsection{Mixing from time-dependent \ $D^0 \to K^0_S\pi^+\pi^-$  \ 
Dalitz plot analysis}

A direct measurement of the mixing  parameters $x$ and $y$ is accomplished 
by a time-dependent amplitude analysis of the self-conjugate decay 
$D^0\to K_S^0 \pi^+\pi^-$. The $D^0$ meson is produced in a well defined 
flavour state. Since the final state is reachable by both $D^0$ and 
$\overline{D}^0$, at any later time one has a mixture of both flavour 
eigenstates. The dominant contribution  to the $D^0$ Dalitz plot is 
the "RS" CF intermediate state $K^{*-}\pi^+$ ($K^{*+}\pi^-$, for the $\overline{D}^0$). There is also a small "WS" $K^{*+}\pi^-$ component,
either from a direct DCS transition or from a net $D^0\to \overline{D}^0$ 
oscillation followed by the CF $\overline{D}^0\to K^{*+}\pi^-$. The  
RS and WS components become time dependent, and this is the mixing signature. 
Assuming no direct \cpv, the mixing parameters $x$ and $y$ are determined 
by a simultaneous fit of the $D^0$ and $\overline{D}^0$ Dalitz plots as a 
function of the decay proper time.

A time-dependent Dalitz plot analysis of the $D^0 \to K^0_S\pi^+\pi^-$
decay was published by Belle\cite{belle.kspipi14},
based on $1.23 \times 10^6$ signal events with a purity of 95.6\%.
The Dalitz plot (DP) distribution is described in terms of 
the two invariants  $s_{\pm} \equiv m^2(K^0_S\pi^{\pm})$.
The decay amplitudes  are parameterized as a sum of quasi-two-body amplitudes,
\begin{equation}
A(s_+,s_-) = \sum a_k e^{i \delta_k} A_k(s_+,s_-),
\end{equation}
for a sample containing only $D^0$ decays, and
\begin{equation}
\overline{A}(s_+,s_-) =  \sum \bar a_k e^{i \bar\delta_k} A_k(s_-,s_+),
\end{equation}
for the sample containing only $\overline{D}^0$ decays. The assumption of no 
direct \cpv imply the equality of the complex coefficients 
$\bar a_k e^{i \bar \delta_k}=a_k e^{i \delta_k}$. The decay model has 
The dependence on $x$ and $y$ appears when the expressions for the decay 
matrix elements are squared.

A Dalitz plot fit is performed assuming no \cp violation, having as free parameters
$x$, $y$, the $D^0$ lifetime, the parameters defining the proper time resolution
function, and the parameters of the decay amplitude model. A fit allowing for
\cpv\ is also performed and includes two more free parameters, 
$|q/p|$ and $\phi=\mathrm{arg}(q/p)$. 

\begin{figure}
\centering

\includegraphics[width=0.3\textwidth]{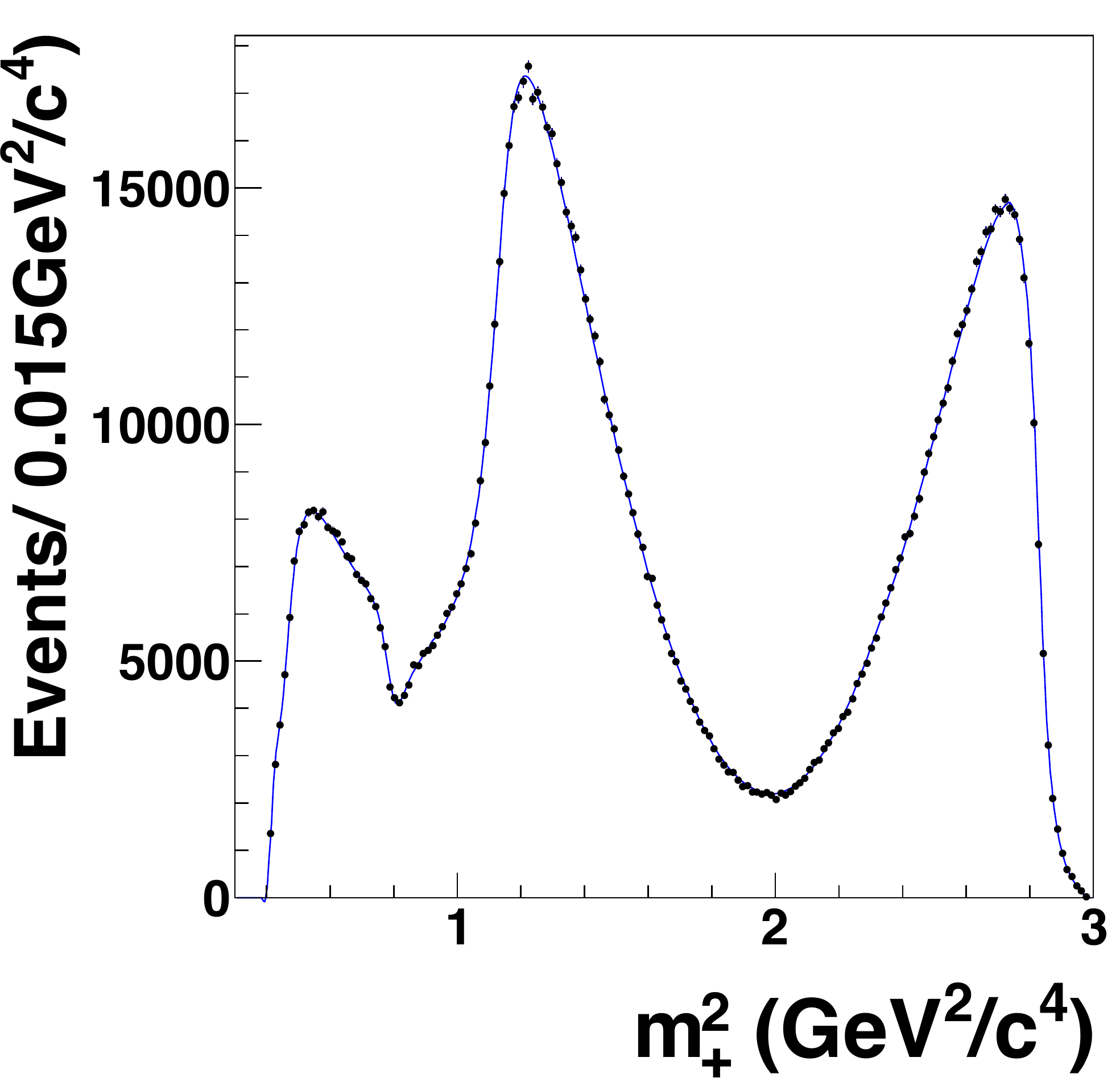}
\includegraphics[width=0.3\textwidth]{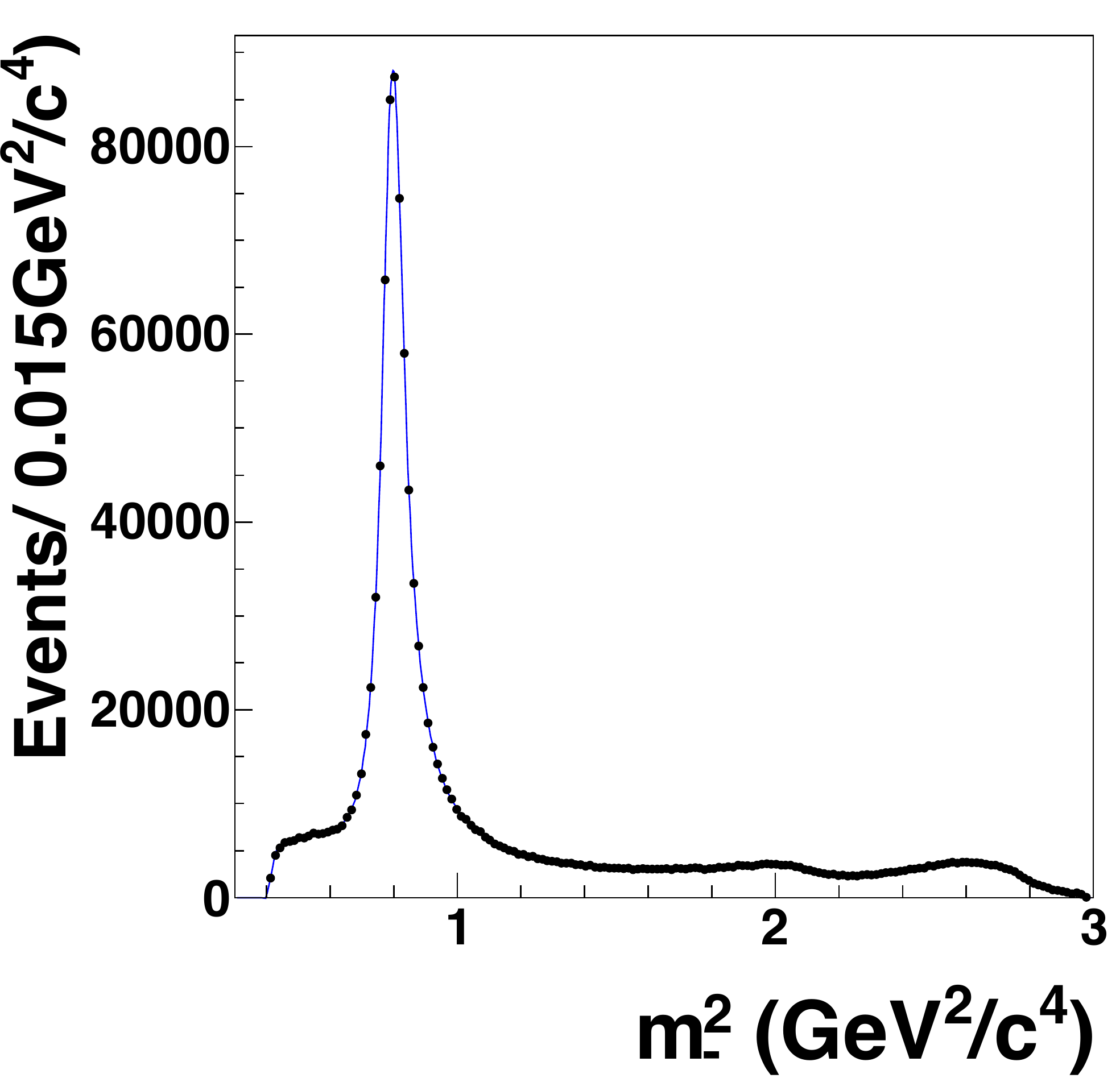}
\includegraphics[width=0.3\textwidth]{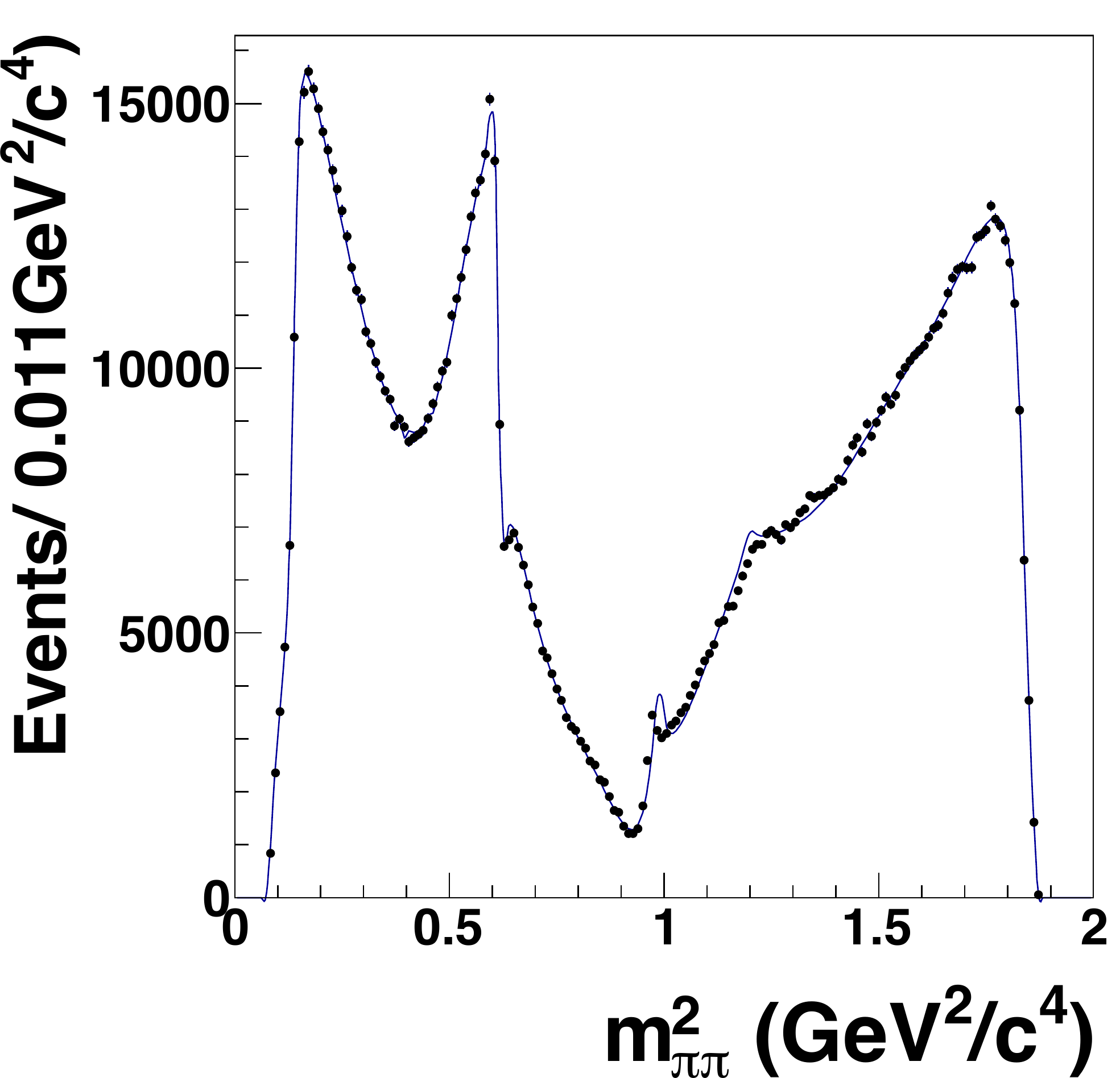}
\caption{Dalitz plot projections onto the $m_+$, $m_-$ and $m_{\pi\pi}$ axes,
from Belle's $D^0 \to K^0_S\pi^-\pi^+$ candidates.}
\label{fig:belle-kspp}
\end{figure}

The  two-body mass projections of the 
$D^0 \to K^0_S\pi^-\pi^+$ candidates from Belle are shown in 
Fig.~\ref{fig:belle-kspp}. The decay model has 14 resonances, a K-matrix 
for $\pi\pi$ S-wave, effective range (LASS) for  $K^0_S\pi$ S-wave 
(40 free parameters), but only an approximate solution found. The uncertainty in
the decay amplitude model is the dominant systematic effect.
The measured values of $x$ and $y$ are shown in Table \ref{tab:kshh}.

\begin{figure}
\centering
\includegraphics[width=0.45\textwidth]{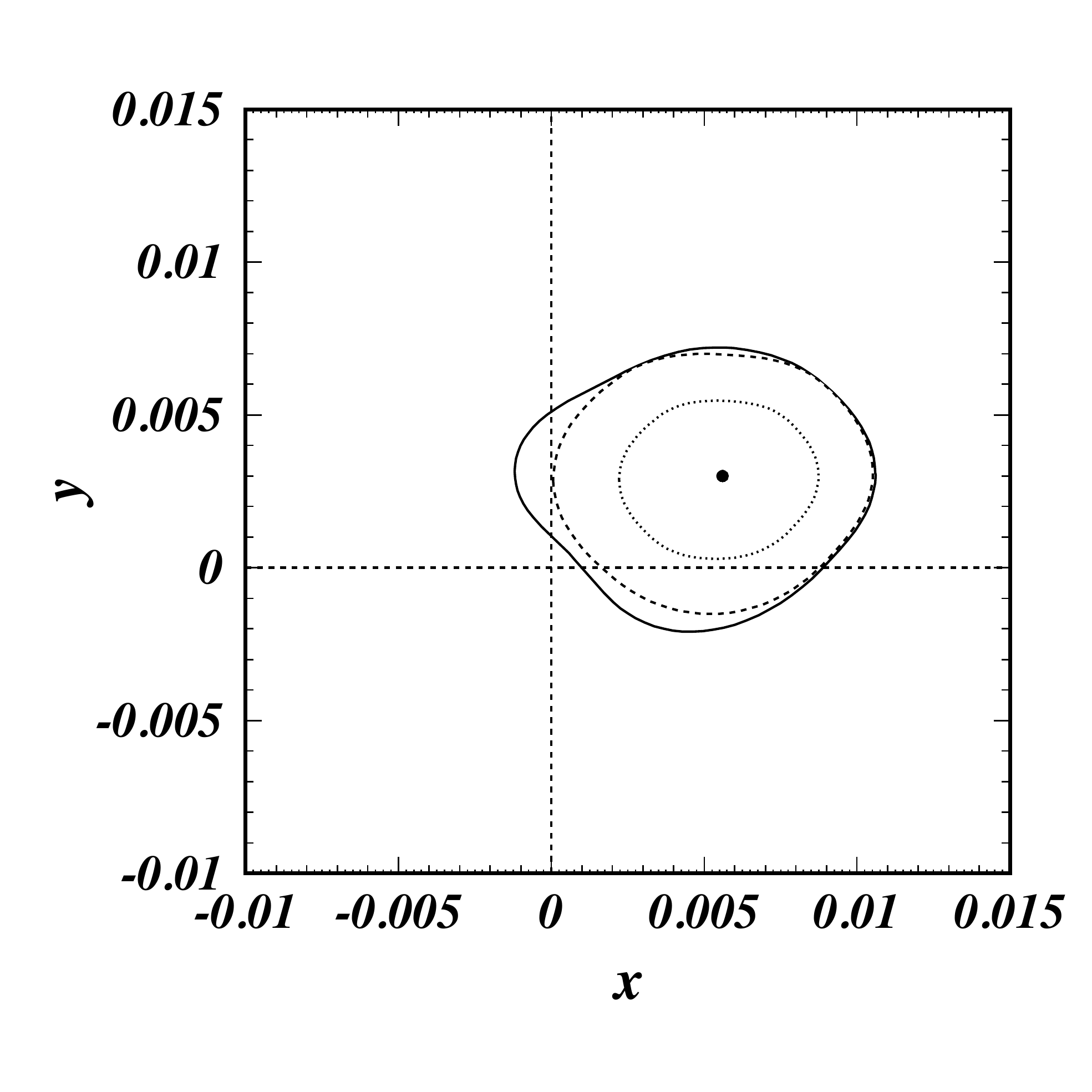}
\caption{Confidence level contours in the $x,y$ plane from Belle fit to the 
$D^0\to K^0_S\pi^+\pi^-$ Dalitz plot. The dotted and dashed lines indicate the
68.3\% and 95\% CL contours from the \cp-conserving fit. The solid line represents
the 95\% CL contour for the \cpv-allowed fit.}
\label{fig:belle-kspp-cl}
\end{figure}

The two-dimensional confidence level contours are shown in Fig. 
\ref{fig:belle-kspp-cl}. 
The significance of \ddbar mixing is estimated to be 2.5 standard deviations 
from the no-mixing point ($x\!=\!y\!=\!0$). In spite of statistical 
significance bellow three standard deviations, this is the most precise
measurement of the  mixing parameters $x$ and $y$. No evidence for \cp\ 
violation was found.

\begin{table}[ht]
\begin{center}
\caption{Measured parameters of \ddbar mixing (\%) from Belle's
time dependent $D^0\to K^0_Sh^+h^-$ Dalitz plot analyses, 
for different hypotheses on \cp symmetry.}
{\begin{tabular}{lc} \\\hline  
Parameter  & Fit result \\
\hline
no {\em CP}V          &                                   \\
$x$(\%) & 0.56 $\pm 0.19^{\ +0.07}_{\ -0.13}$  \\ 
$y$(\%) & 0.30 $\pm 0.15^{\ +0.05}_{\ -0.08}$  \\
\\
{\em CP}V allowed &   \\
$x$(\%)   &  0.56 $\pm 0.19^{+0.07}_{-0.11}$  \\ 
$y$(\%)   &  0.30 $\pm 0.15^{+0.05}_{-0.09}$         \\
$|q/p|$  &  0.90$^{+0.16 \ +0.08}_{-0.15 \ -0.07}$       \\
$\mathrm{arg}(q/p)(^{\circ})$   &  -6 $\pm$ 11 $\pm 3^{\ +3}_{\ -4}$      \\
\hline      
\end{tabular}
\protect\label{tab:kshh}}
\end{center}
\end{table}

\newpage

\section{CP violation searches in time-integrated rates}

\subsection{$\Delta A_{CP}$ \  from \ $\overline{B} \to D^0\mu^- X$}

The time-integrated \cp asymmetry in  $D^0$  decays to a $CP$ eigenstate  
$f$  ($f\!=\!K\overline{K},\pi\pi$) is mostly a measurement of direct 
\cpv,

\[ 
A_{CP}(f)\equiv \frac{\Gamma(D^0\to f)-\Gamma(\overline{D}^0\to f)}
{\Gamma(D^0\to f)+\Gamma(\overline{D}^0\to f)} \ \simeq \ 
a_{CP}^{\mathrm{dir}}(f) + \frac{\langle t\rangle}{\tau}a_{CP}^{\mathrm{ind}},
\]

The term $a_{CP}^{\mathrm{ind}}$ represents \cpv\  in mixing and/or in 
interference  between mixing and decay. It is universal, to 
a  good approximation, with a contribution that  depends on the experimental 
decay-time acceptance. 

In the SM, direct \cp\ asymmetry in $K^-K^+$  and  $\pi^-\pi^+$ final states
are expected to have opposite signs, so measuring their difference
$\Delta A_{CP}  \equiv \ A_{CP}(K^-K^+) - A_{CP}(\pi^-\pi^+)$,  one benefits 
not only from from a higher sensitivity to \cpv\ and from the cancellation of
systematics and production and detection asymmetries.

\begin{figure}
\centering
\includegraphics[width=0.5\textwidth]{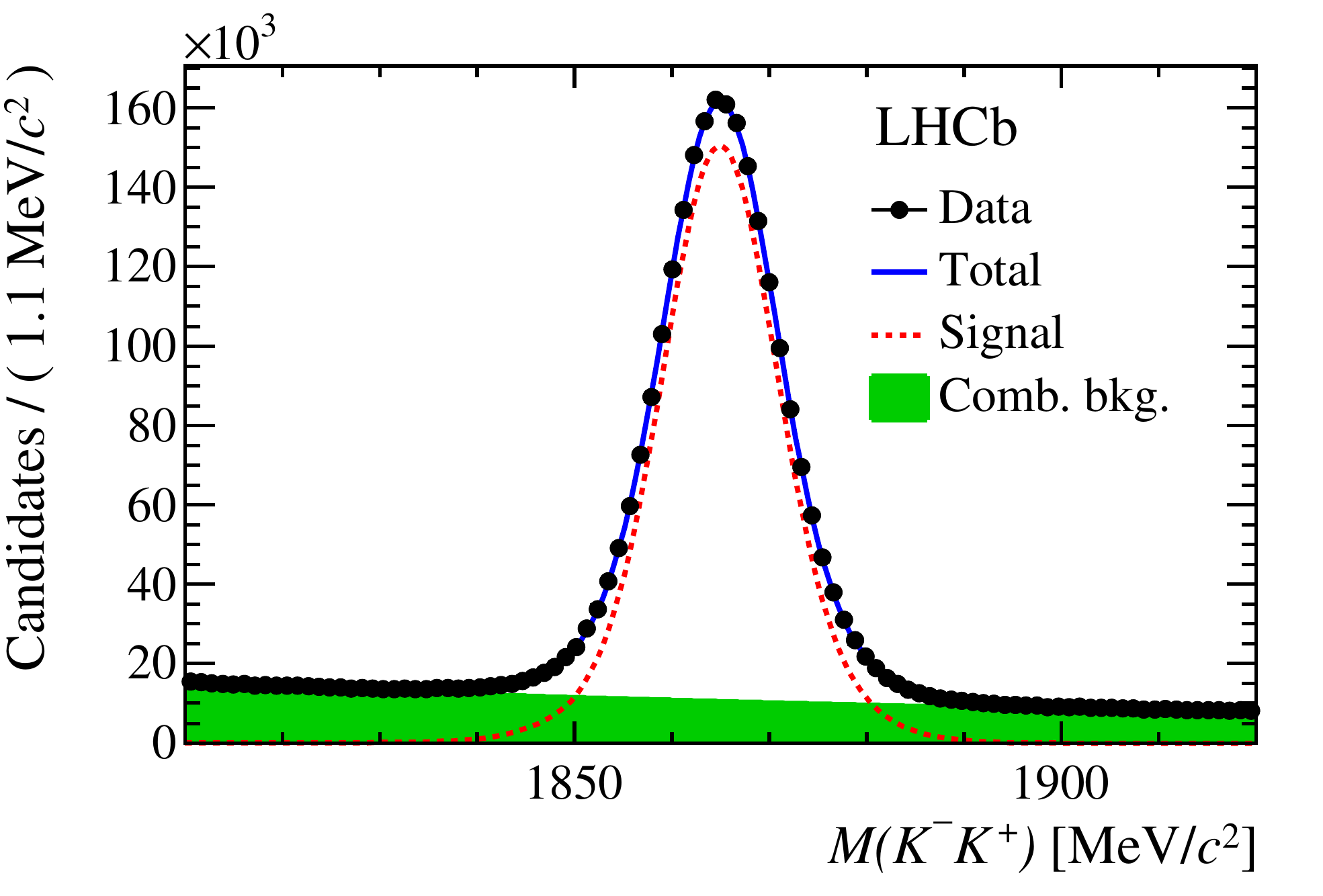}\includegraphics[width=0.5\textwidth]{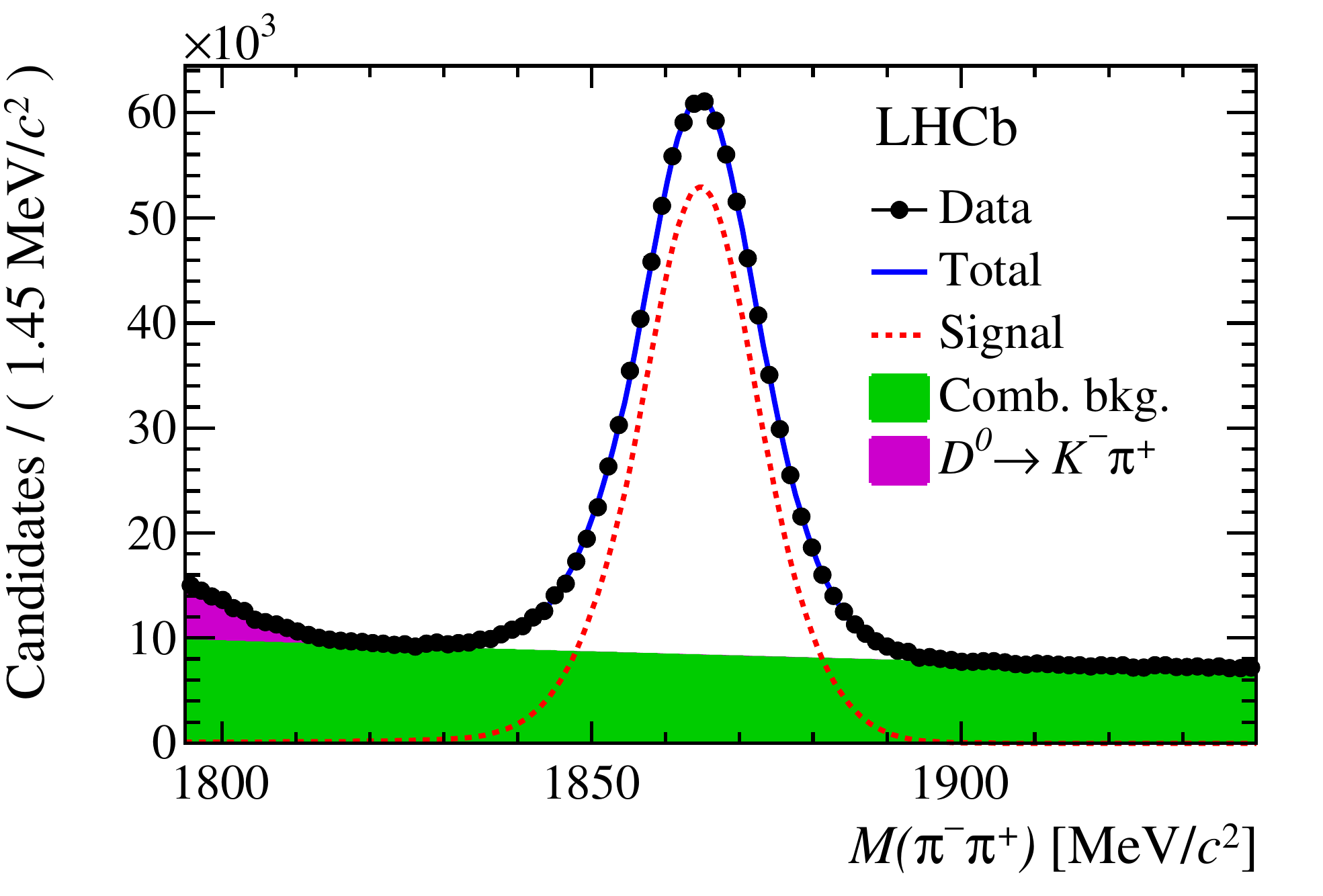}
\caption{Invariant mass distributions of muon-tagged $D^0\to K^-K^+$ (left)
and $D^0\to \pi^-\pi^+$ (right), from LHCb, with fit results superimposed. }
\label{fig:semilep-mass}
\end{figure}

LHCb released a new $\Delta A_{CP}$ \ measurement\cite{lhcb-dacp-semi2} 
using $D^0$ decays from partially reconstructed \ $\overline{B} \to D^0\mu^- X$
decays. The full run I data set (3 fb$^{-1}$) contains 
$1.99\times 10^{6}$ and $0.78\times 10^{6}$ $K^-K^+$ and $\pi^-\pi^+$ 
candidates.  The LHCb signals are shown in Fig.~\ref{fig:semilep-mass}.
These events, with the $D^0$ flavour tagged by the muon sign, form an 
independent set from that containing  $D^0$ directly produced in the primary 
$pp$ interaction and tagged by the pion sign in $D^{*+} \to D^0\pi^+$.

LHCb measures the total, or {\em raw} charge asymmetry,
\[
A_{\mathrm{raw}}=\frac{\Gamma(D^0\to f)\varepsilon(\mu^-)\mathcal{P}(D^0)-
\Gamma(\overline{D}^0\to f)\varepsilon(\mu^+)\mathcal{P}(\overline{D}^0)}
{\Gamma(D^0\to f)\varepsilon(\mu^-)\mathcal{P}(D^0)+
\Gamma(\overline{D}^0\to f)\varepsilon(\mu^+)\mathcal{P}(\overline{D}^0)} \ \simeq \
A_{CP}^f+A^{\mu}_D + A^B_{\mathcal{P}}.
\]

The detection and production asymmetries are, to first order, independent of
the $D^0$ decay and cancel in the difference between the $K^-K^+$ and $\pi^-\pi^+$ 
raw asymmetries,
\[
\Delta A_{CP} = A_{\mathrm{raw}}(K^-K^+) - A_{\mathrm{raw}}(\pi^-\pi^+) \simeq
A_{CP}(K^-K^+) - A_{CP}(\pi^-\pi^+)
\]

Second order effects are studied and accounted by as a systematic uncertainty.
Using the Cabibbo favoured decays \ $D^0\to K^-\pi^+$, 
\ $D^+\to K^-\pi^+ \pi^+$ \ and \ $D^+\to K^0_S\pi^+$, the asymmetries \ 
$A^{\mu}_D$ and $A^B_{\mathcal{P}}$ are determined, allowing the measurement 
of the individual asymmetries:
\[
A_{\mathrm{raw}}(K^-\pi^+) = A_D^{\mu} - A_{\mathcal{P}}^B - A_D(K^-\pi^+),
\]
\[
A_{CP}(K^-K^+) = A_{\mathrm{raw}}(K^-K^+) - A_{\mathrm{raw}}(K^-\pi^+) -
A_D(K^-\pi^+)
\]

The results from the updated LHCb measurement are
\begin{eqnarray}
\Delta A_{CP} =& (0.14 \pm 0.16 \pm 0.08)\%, \nonumber \\
A_{CP}(KK) =& (-0.06 \pm 0.15 \pm 0.010)\%, \nonumber \\
A_{CP}(\pi\pi) =& (-0.20 \pm 0.15 \pm 0.10)\%, 
\end{eqnarray}

\begin{figure}
\centering
\includegraphics[width=0.45\textwidth]{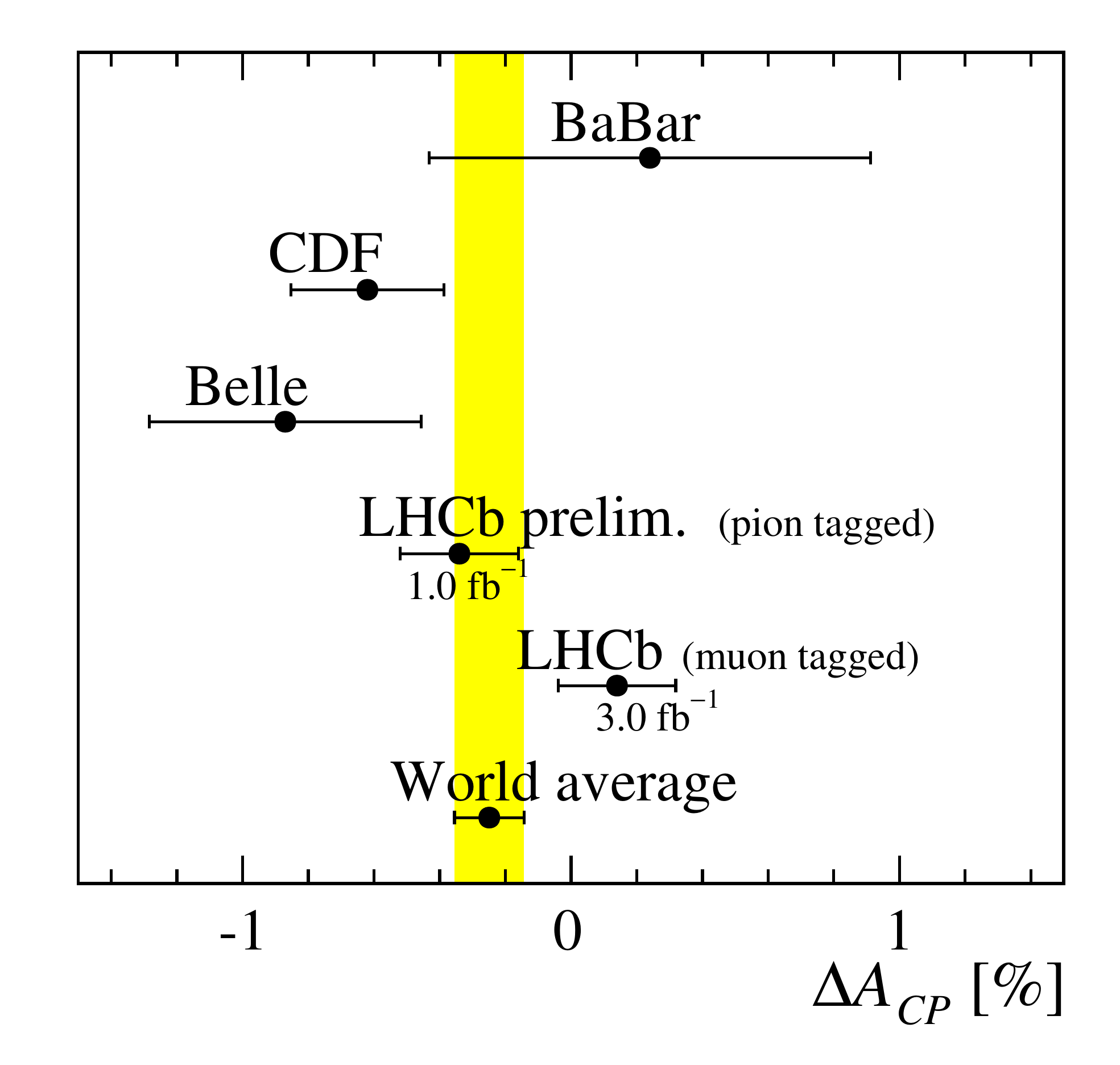}\includegraphics[width=0.45\textwidth]{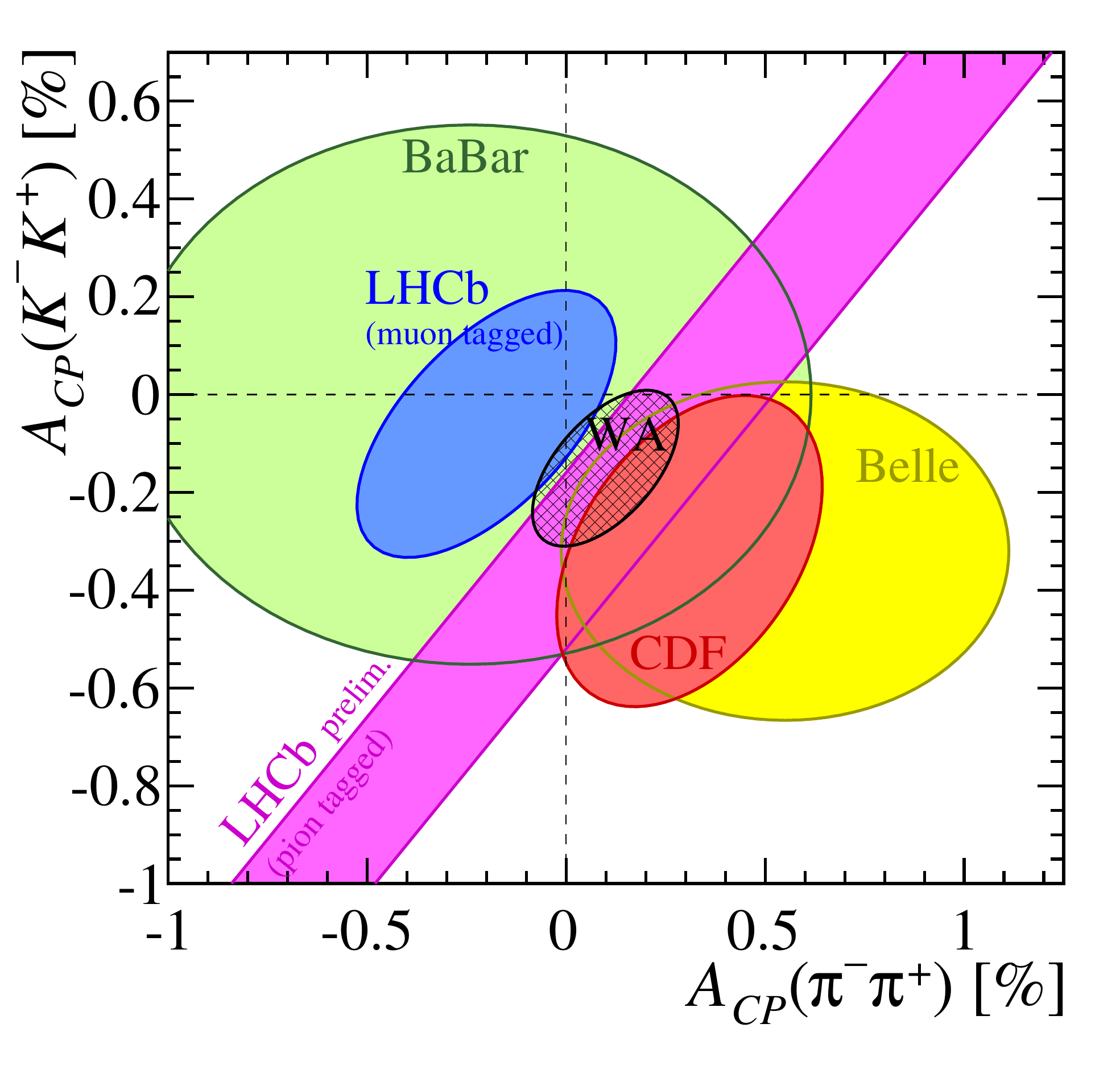}
\caption{Overview of \cp violation measurements in $D^0\to K^-K^+$
and $D^0\to \pi^-\pi^+$ decays from \cite{babar.dacp}, 
\cite{lhcb-dacp-semi2}, \cite{lhcb.dacp.2013}, \cite{cdf.dacp}, and
\cite{belle.dacp}. On the left, measurements of the  difference, $\Delta A_{CP}$ are shown together with 
the world average. On the right, the  $A_{CP}(KK)$ versus $A_{CP}(\pi\pi)$ 
plane with  the $68\%$ confidence level contours is displayed. 
The new world averages are obtained neglecting any effect from
indirect \cp violation. Reproduced from the supplementary material of
Ref.~\cite{lhcb-dacp-semi2}.}
\label{fig:new-wa}
\end{figure}

No evidence for \cp violation is found in this measurement. A  world 
average of $\Delta A_{CP}$ and of the individual asymmetries
was computed by the authors of Ref.~\cite{lhcb-dacp-semi2}. 
An overview of the various measurements of $\Delta A_{CP}$ is shown in
the plot on the left in Fig.~\ref{fig:new-wa}. A world average,
$\Delta A_{CP}=(-0.25\pm0.11)\%$ is obtained neglecting indirect \cp 
violation effects. In the right plot of Fig.~\ref{fig:new-wa} an overview
of the existing $A_{CP}(KK)$ and $A_{CP}(\pi\pi)$ are shown.
World averages are found to be $A_{CP}(KK)=(-0.15\pm0.11)\%$ and 
$A_{CP}(\pi\pi)=(0.10\pm0.12)\%$ with a correlation $\rho=0.57$.

\subsection{$D^+_{(s)} \to K_S^0h^+$}

Direct \cpv\ searches in charged $D$ mesons are a natural complement to the 
measurements performed with neutral $D$. In particular, the modes 
$D^+\to K^0_SK^+$\  and \ $D^+_s \to K^0_S\pi^+$ have very similar
amplitudes as those for $D^0 \to K^+K^-$ \ and \ $D^0 \to \pi^+\pi^-$.

\begin{figure}
\includegraphics[width=0.45\textwidth]{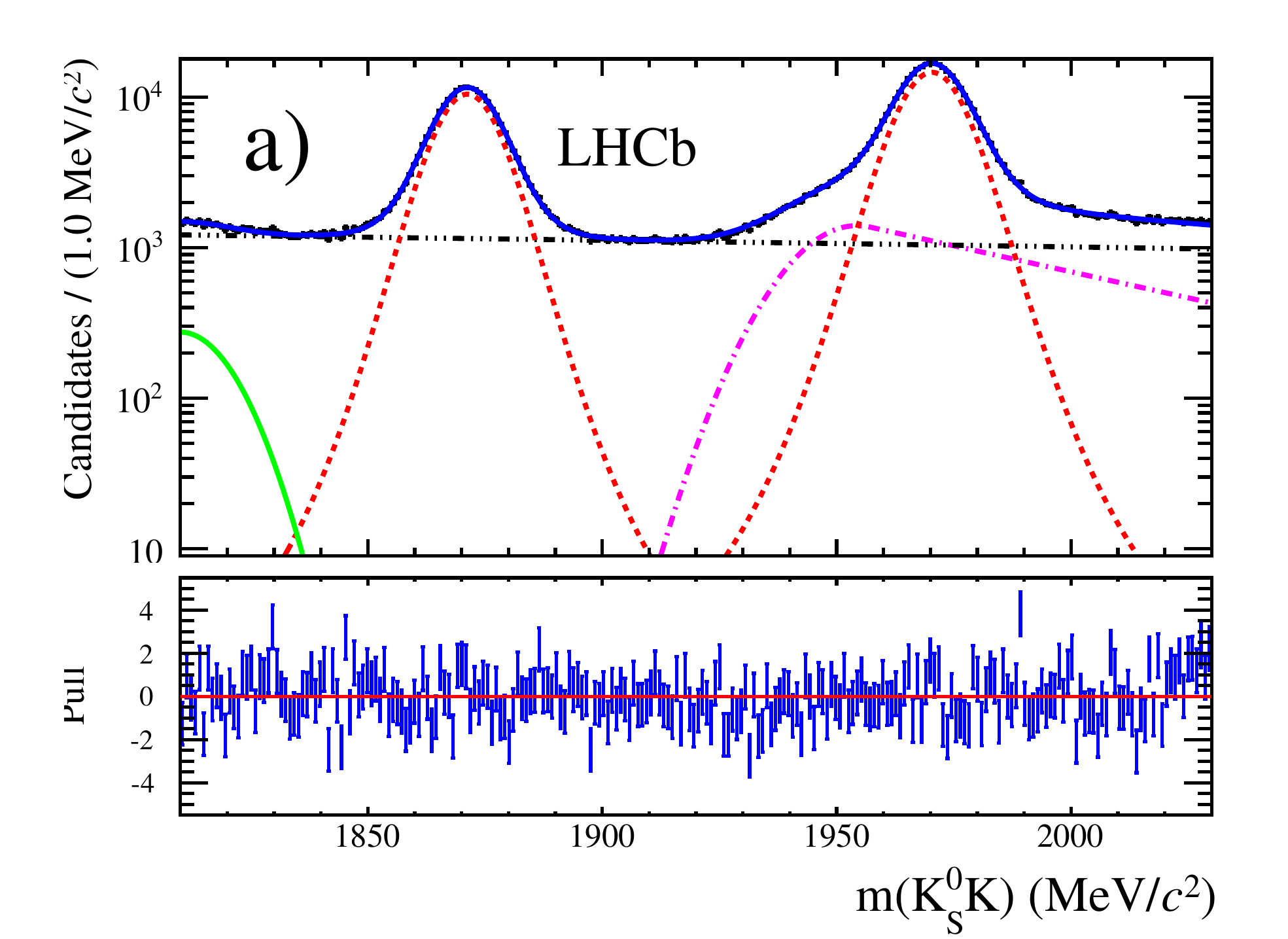}\includegraphics[width=0.45\textwidth]{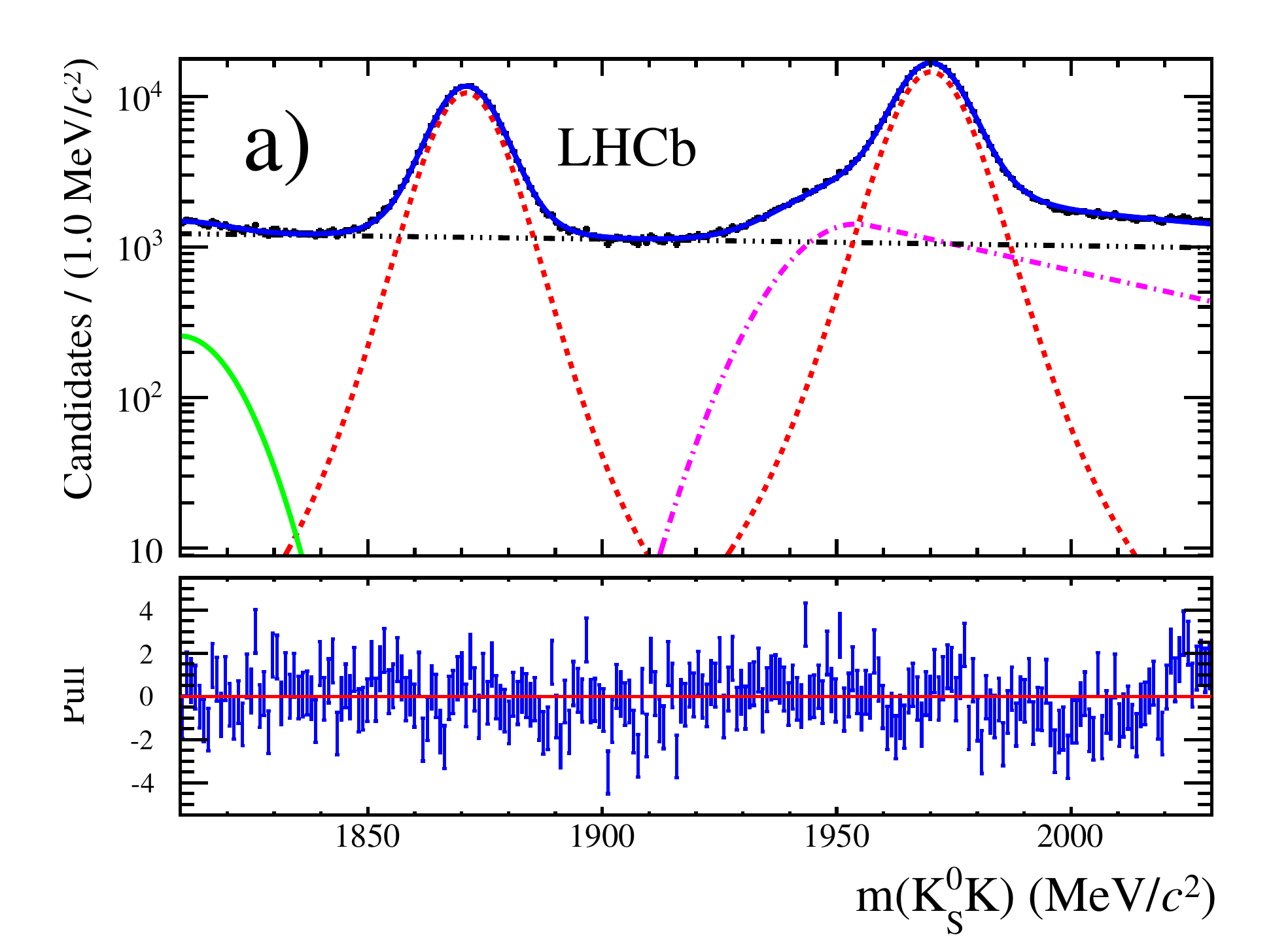}
\includegraphics[width=0.45\textwidth]{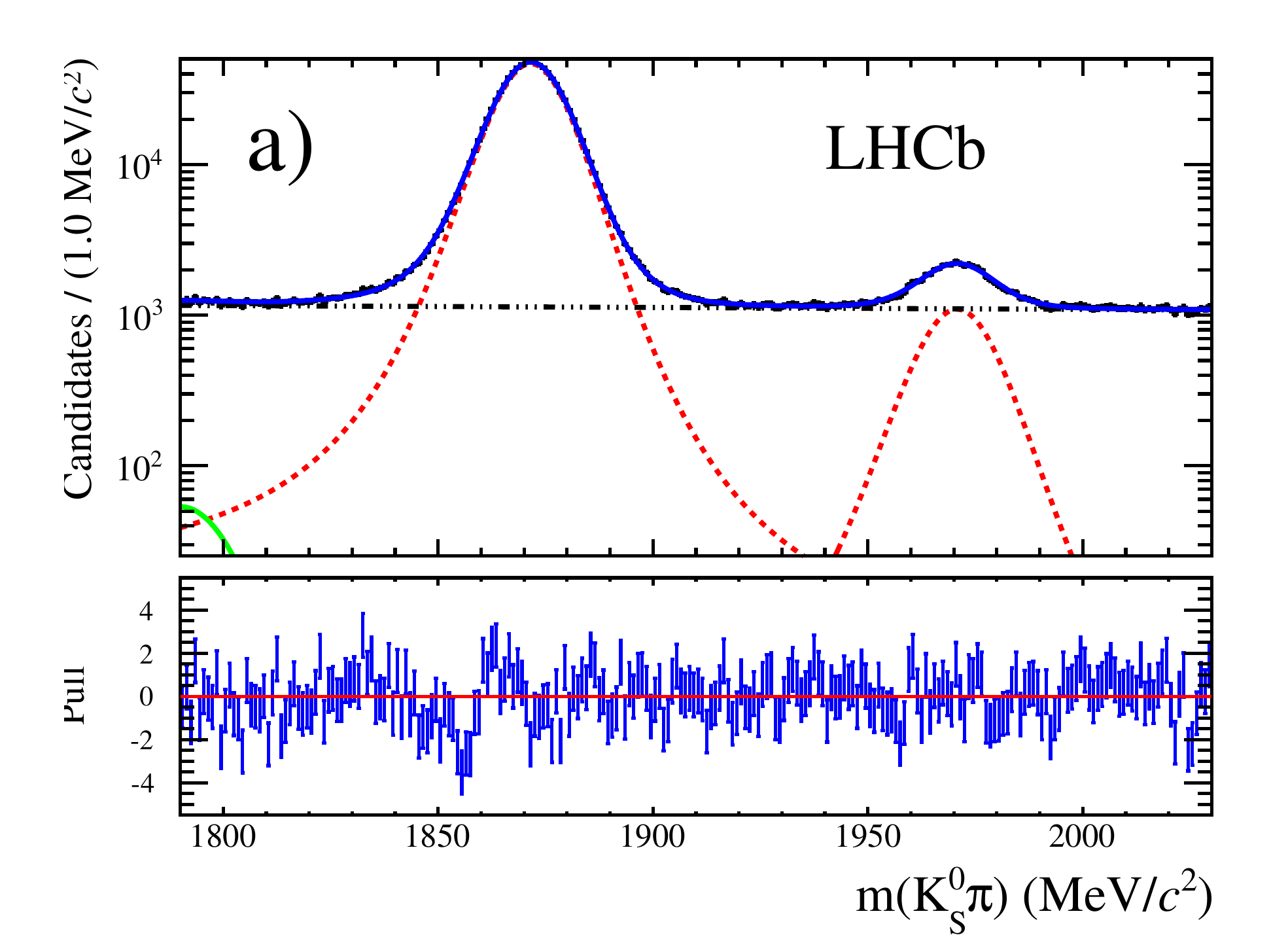}\includegraphics[width=0.45\textwidth]{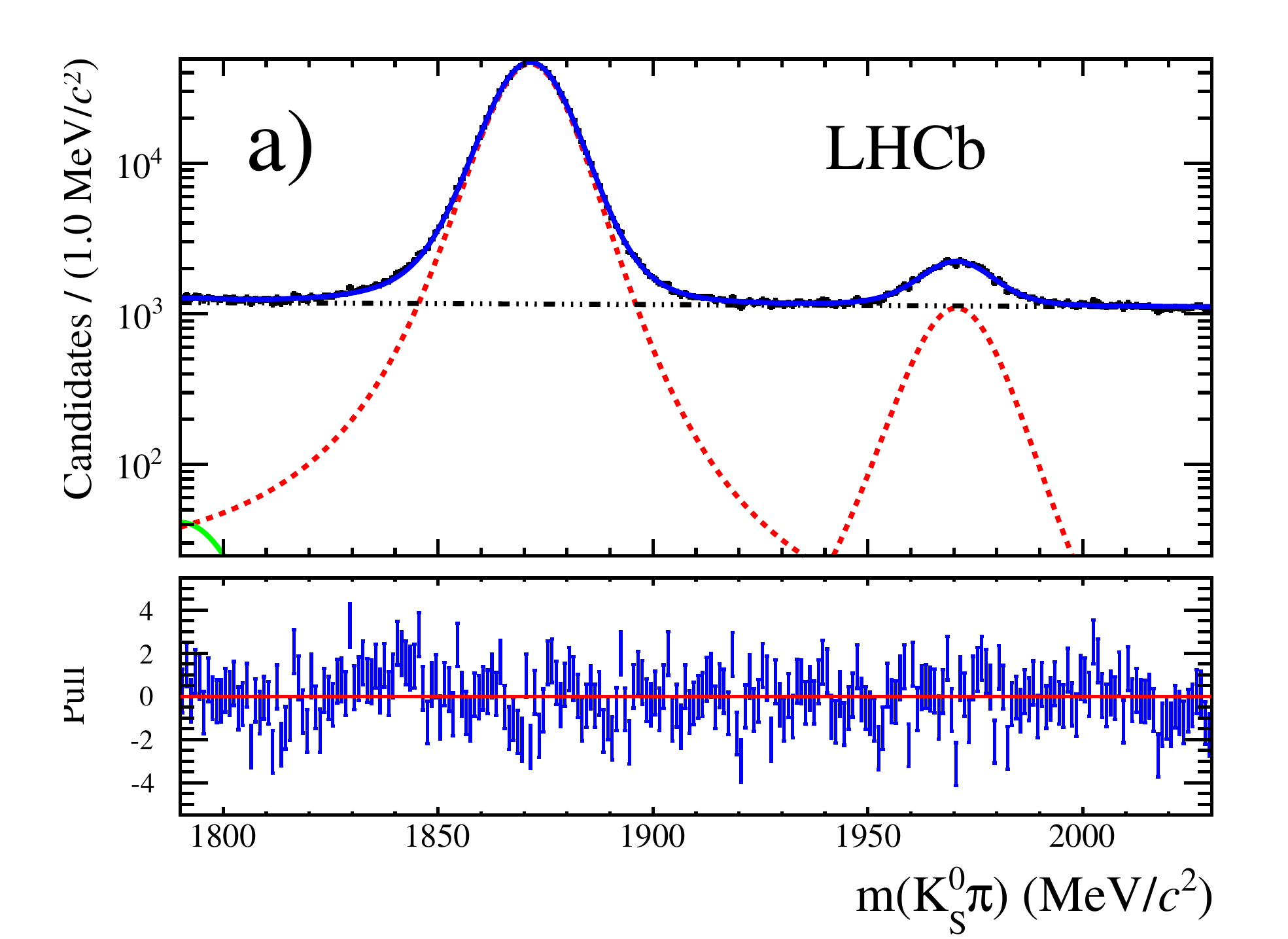}
\caption{Invariant mass distributions for the $D^+_{(s)}\to K^0_SK^+$ (top left), $D^-_{(s)}\to K^0_SK^-$ (top right),
$D^+_{(s)}\to K^0_S\pi^+$ (bottom left) and$D^-_{(s)}\to K^0_S\pi^-$ (bottom right), from LHCb. The fit result is 
overlaid, showing the contribution from the different backgrounds.}
\label{fig:ksk}
\end{figure}

LHCb measures the \cp\ asymmetry in $D^+\to K^0_SK^+$ and $D^+_s\to K^0_S\pi^+$
using the full 3 fb$^{-1}$ data set from run I (2011+2012)\cite{lhcb.ksh}.
The signal yields are $1.0\times 10^6$
$D^+\to K^0_SK^+$ and $1.2\times 10^5$ $D^+_s\to K^0_S\pi^+$ decays. The 
analysis uses only  $K_S^0$  that decay at the vertex detector. Since these 
are very short lived $K_S^0$, the contribution to the measured charge asymmetry 
from \cp violation in $K^0-\overline{K}^0$ mixing is found to be negligible. 
In Fig. \ref{fig:ksk}, the invariant mass distributions of the selected 
candidates are shown.

The observed charge asymmetry is a sum of the physical \cp asymmetry plus contributions from production and detection asymmetries,
\[
\mathcal{A}_{\mathrm{raw}}^{D^+_{(s)}\to K_S^0h^+}\!\approx \mathcal{A}_{CP}^{D^+_{(s)}\to K_S^0h^+}\!+
\mathcal{A}_{\mathrm{prod}}^{D^+_{(s)}} + \mathcal{A}_{\mathrm{det}}^{h^+} +  
\mathcal{A}_{K^0},
\]

The two observables, $\mathcal{A}_{CP}^{D^+\to K_S^0K^+}$ and
$\mathcal{A}_{CP}^{D^+_s\to K_S^0\pi^+}$ are obtained from five measurements, 
namely the charge asymmetries of the four Cabibbo favoured decays 
$D^+_{(s)}\to K_S^0h^+$ ($h=K,\pi$) plus the asymmetry of $D^+_s\to \phi\pi^+$,
used as a control channel:
\begin{eqnarray}
\mathcal{A}_{CP}^{D^+\to K_S^0K^+}\!&\approx& \left [\mathcal{A}_{\mathrm{meas}}^{D^+\to K_S^0K^+}-
\mathcal{A}_{\mathrm{meas}}^{D^+_s\to K_S^0K^+}\right ] - 
\left [\mathcal{A}_{\mathrm{meas}}^{D^+\to K_S^0\pi^+}-
\mathcal{A}_{\mathrm{meas}}^{D^+_s\to \phi\pi^+}\right ]-\mathcal{A}_{K^0},  \nonumber \\
\mathcal{A}_{CP}^{D^+_s\to K_S^0\pi^+}\!&\approx& \mathcal{A}_{\mathrm{meas}}^{D^+_s\to K_S^0\pi^+} -
\mathcal{A}_{\mathrm{meas}}^{D^+_s\to \phi\pi^+}-\mathcal{A}_{K^0}. \nonumber
\end{eqnarray}

Only $K^0_S$ with short decay time are used, the contribution from 
\cpv \ and \ $K^0\!-\!\overline{K}^0$ mixing is negligible. The value for 
$\mathcal{A}_{K^0}$ is (0.7$\pm$0.2)\%, and is
dominated by the difference between $K^0$ and $\overline{K}^0$ in interaction 
with the detector material, $\sigma(K^0N)\neq \sigma(\overline{K}^0N)$.

The cancellation of the production and detection asymmetries is ensured by a
event weighting procedure, used to equalize small differences in kinematic distributions of the final states. In addition, fiducial cuts remove a small 
fraction of events with large raw asymmetries in the momentum space. In these
events the lowest momentum particle tends to be deflected  out of the detector.

The values of $A_{CP}(D^+\to K^0_SK^+)$ and $A_{CP}(D^+_s\to K^0_S\pi^+)$
are consistent with zero: 

\begin{eqnarray}
A_{CP}^{D^+\to K_S^0K^+} &= (+0.03\pm0.17\pm0.14)\% \nonumber \\
A_{CP}^{D^+_s\to K_S^0\pi^+} &= (+0.38\pm0.46\pm0.17)\%.  \nonumber \\           
\end{eqnarray}

\subsection{$D^+ \to \pi^-\pi^+\pi^+$}

Three-body decays offer unique opportunities in \cpv\ searches:

\begin{itemize}
\item asymmetries localized in specific regions of the Dalitz may be 
significantly larger than phase-space integrated ones, e.g. large 
asymmetries observed in charmless $B^+\to h_1^-h_2^+h_3^+$  
(see I. Bediaga's talk in this conference);
\item local \cp \ asymmetries may change sign across the phase space;
\item the pattern of local \cp\ asymmetries brings additional information on
the underlying dynamics (other than a single number).
\end{itemize}

The main strategy in the search for \cpv\ in three-body decays of $D$ mesons
is to perform a direct, model-independent comparison between \ $D^+_{(s)}$ \ 
and \ $D^-_{(s)}$ Dalitz plots. The comparison can be made either using a 
binned~\cite{anis} or an unbinned~\cite{mwillians} approach. The drawback to 
this approach is that, in the absence of a signal, one cannot set limits on 
\cpv\ effects. In the event of a \cpv\ signal, this procedure
should be followed by a full amplitude analysis.

The binned technique is being largely utilized to search for localized \cp\
asymmetries. The combined $D^{\pm}_{(s)}$ Dalitz plot is divided into bins. For 
each the statistical significance of the asymmetry
\begin{equation}
S^i_{CP}= \frac{N_i(D^+)-\alpha N_i(D^-)}{\sqrt{\alpha [\sigma^2_i(D^+)+ \sigma^2_i(D^-)]}}
\end{equation}
is calculated, where $N_i(D^+)$ and $N_i(D^-)$ are the number of $D^+$ and
$D^-$ decays in each bin $i$. The uncertainties $\sigma_i(D^+)$ and
$\sigma_i(D^-)$ are usually taken as $\sqrt{N_i(D^+)}$ and $\sqrt{N_i(D^-)}$. 
The parameter $\alpha$ is the ratio between the total number of $D^+$ and $D^-$ 
events and is used as a correction due to a global production asymmetry. 

The distribution of $S^i_{CP}$ is normal under the hypothesis of \cp conservation. 
A $\chi^2$ test using $\chi^2=\sum^{N}_{i=1}S^i_{CP}$ provides a numerical
evaluation for the degree of confidence for the assumption that the differences between the $D^+$ and $D^-$ Dalitz plots are driven only by statistical fluctuations. 

\begin{figure}
\centering
\includegraphics[width=6.5cm]{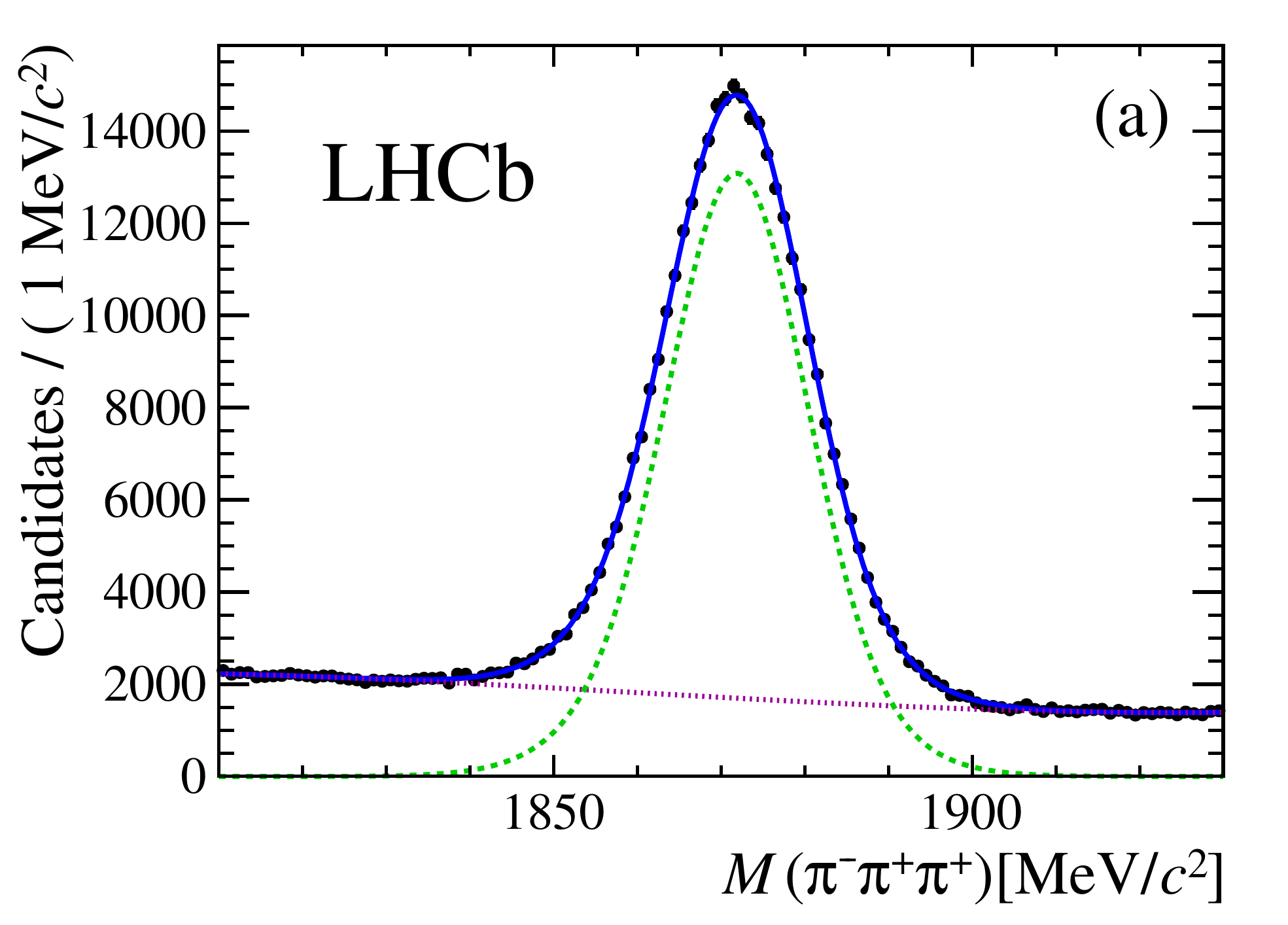}
\includegraphics[width=6.5cm]{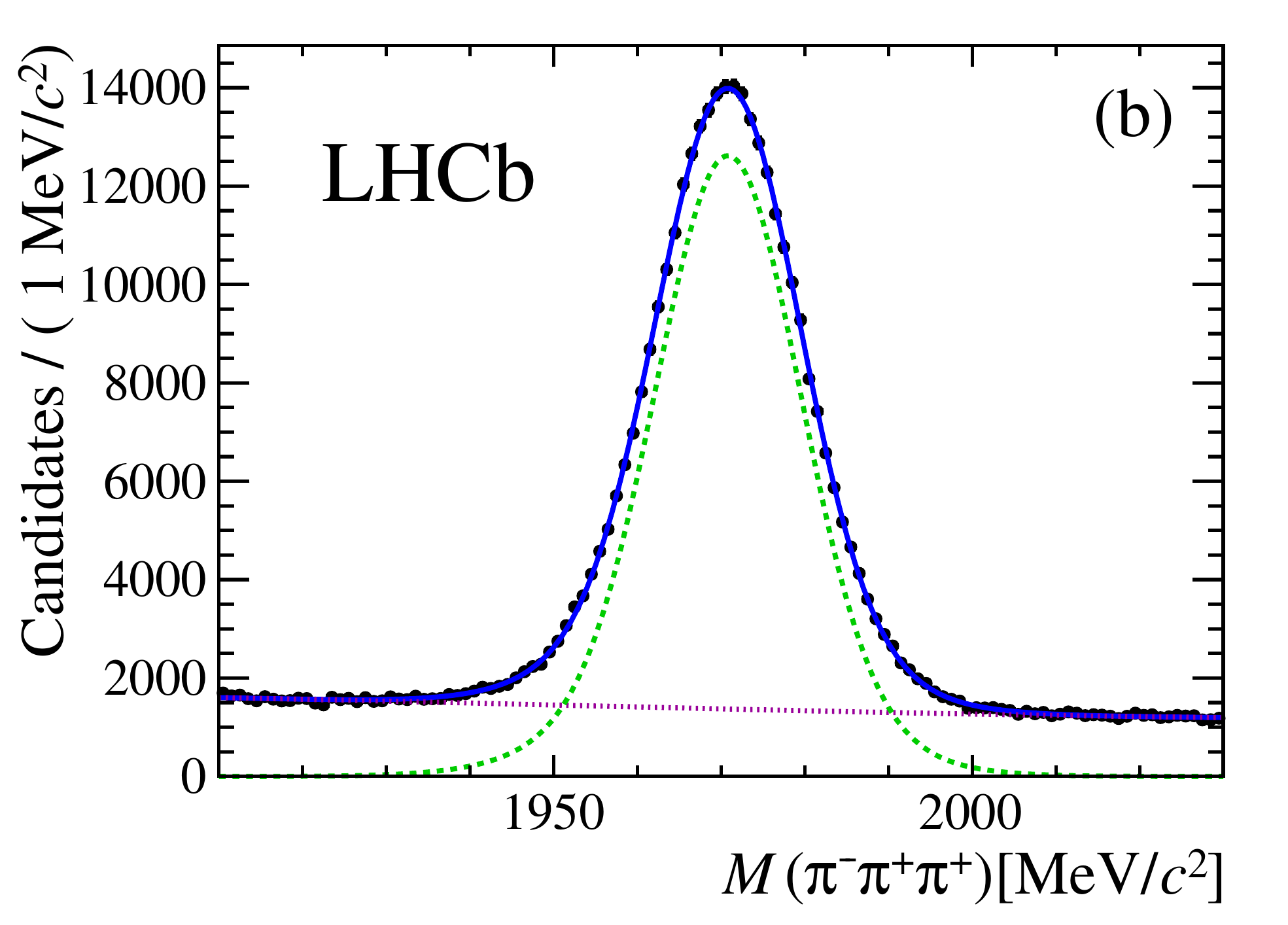}

\includegraphics[width=6.5cm]{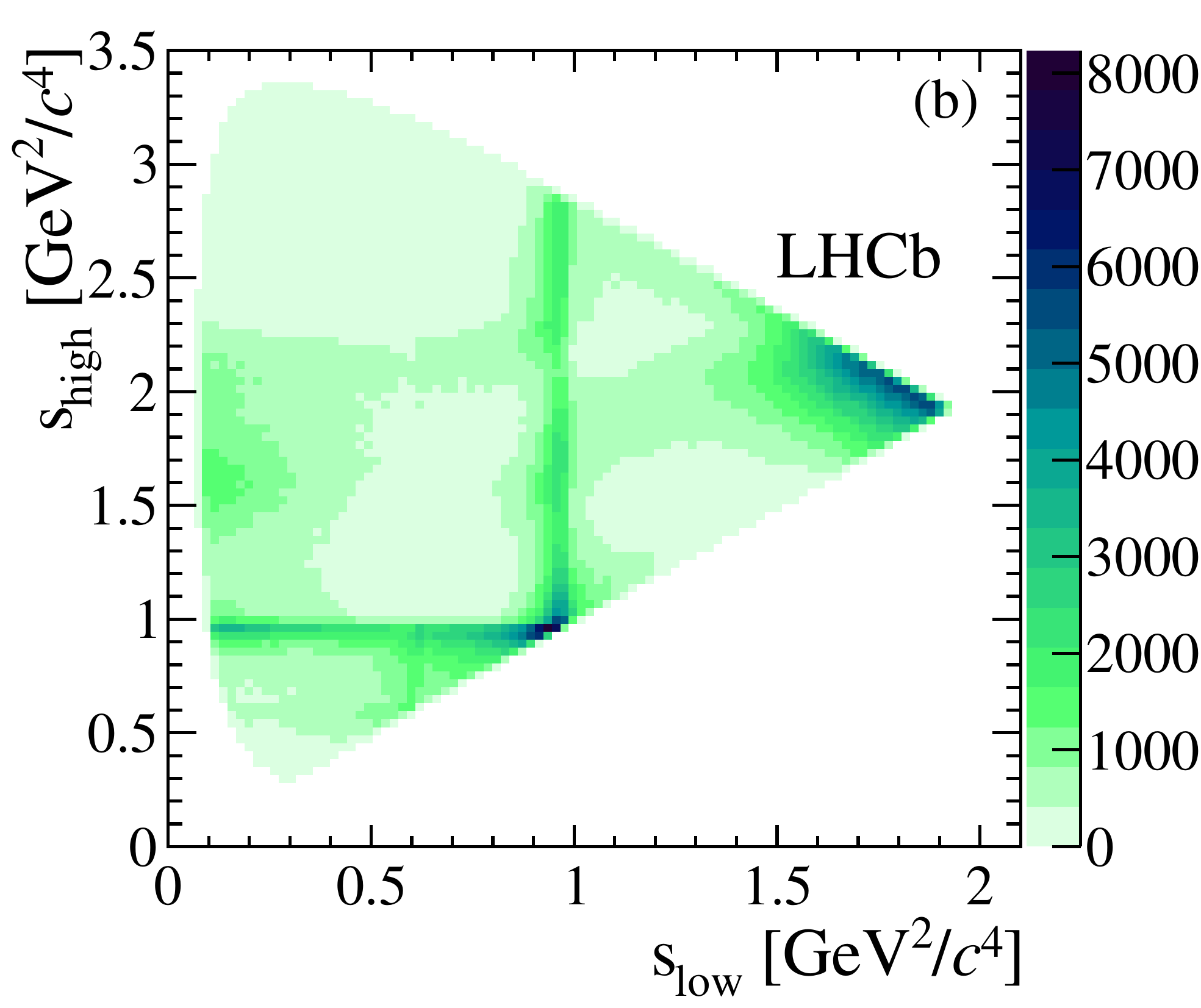}
\includegraphics[width=6.5cm]{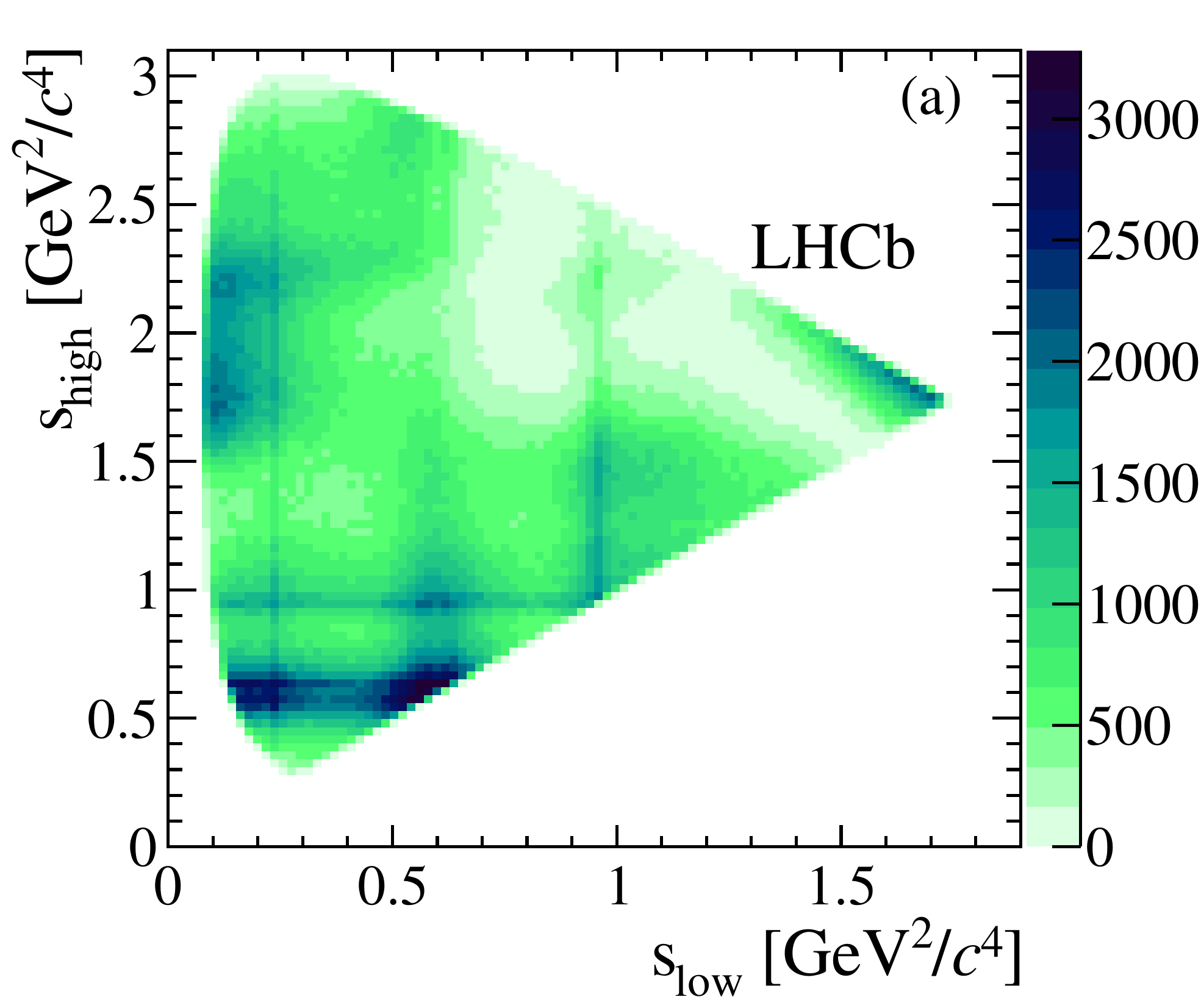}
\caption{The $D^+, D_s^+ \to \pi^-\pi^+\pi^+$ signals and the 
corresponding Dalitz plots.}
\label{fig:ppp}
\end{figure}

LHCb used this model independent approach to look for a \cpv\ signal in the 
Cabibbo suppressed decay $D^+\to \pi^-\pi^+\pi^+$, using the 1 fb$^{-1}$ of 
data collected  2011\cite{lhcb.dppp}. The $D^+\to \pi^-\pi^+\pi^+$ and
$D^+_s \to \pi^-\pi^+\pi^+$ signals and the corresponding Dalitz plots are
shown in Fig. \ref{fig:ppp}. There are approximately 2.68 and 2.70 $\times$
$10^6$ $D^+$ and $D^+_s$ decays, respectively.

The absence of localized asymmetries arising from instrumental and/or 
production effects are tested with the CF decay $D^+_s\to \pi^-\pi^+\pi^+$.  
Inspection on the signal sidebands show that the background is also
insensitive to any possible instrumental asymmetries. The anisotropy method 
is applied to $\pi^-\pi^+\pi^+$ candidates with invariant mass within a 
two $\sigma$ interval around the $D^+$ mass.

\begin{figure}
\centering
\includegraphics[width=6.5cm]{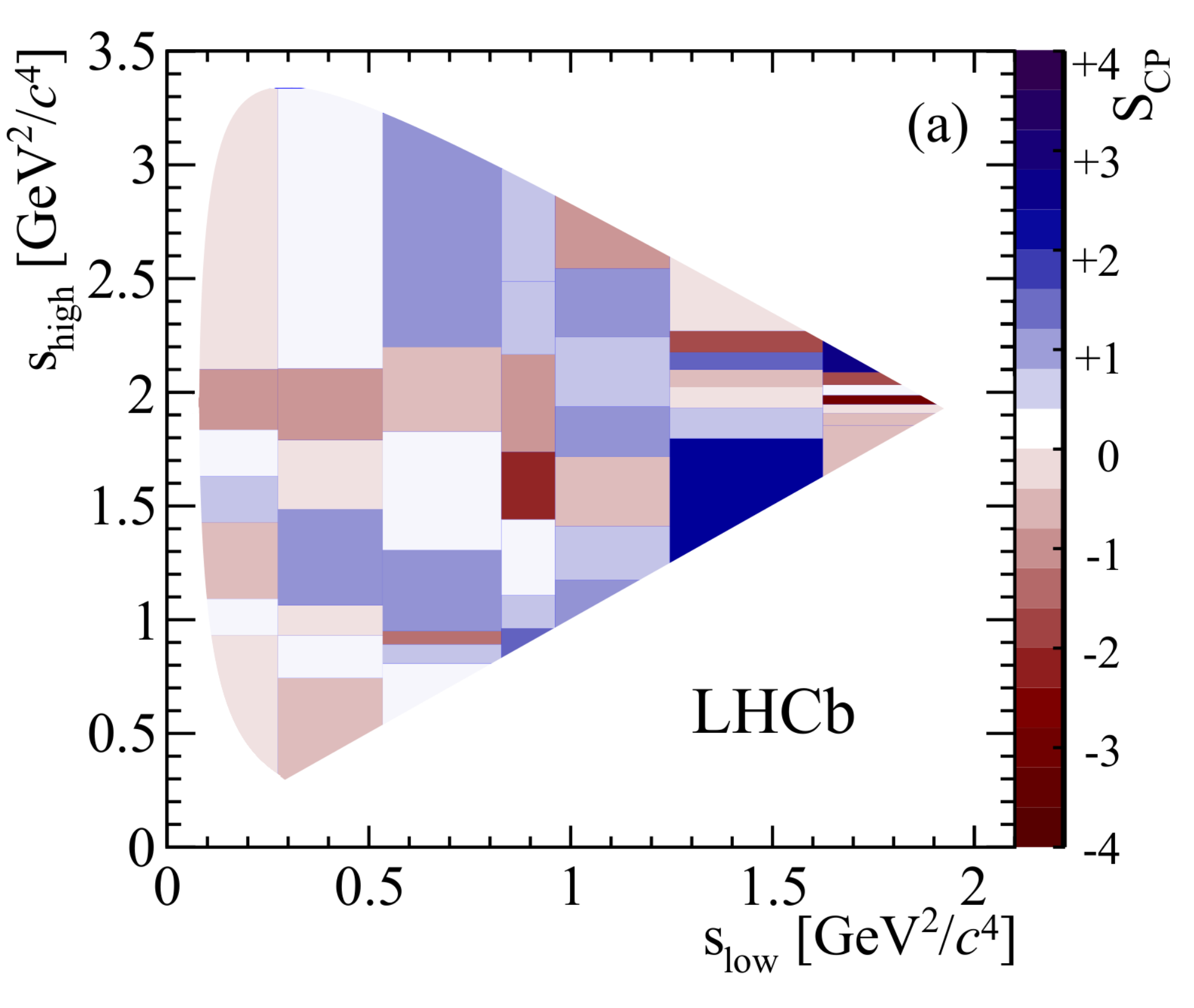}
\includegraphics[width=6.5cm]{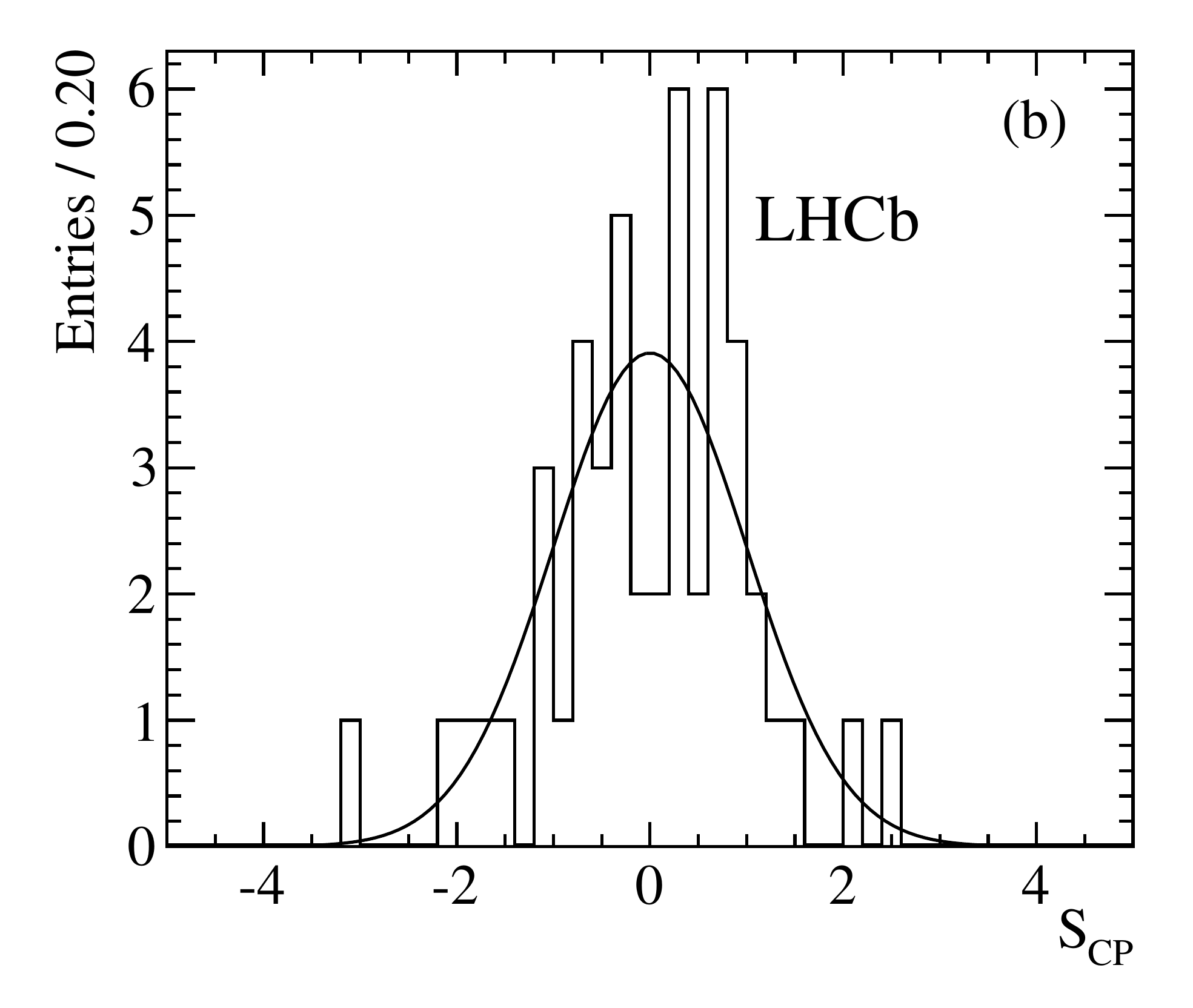}

\includegraphics[width=6.5cm]{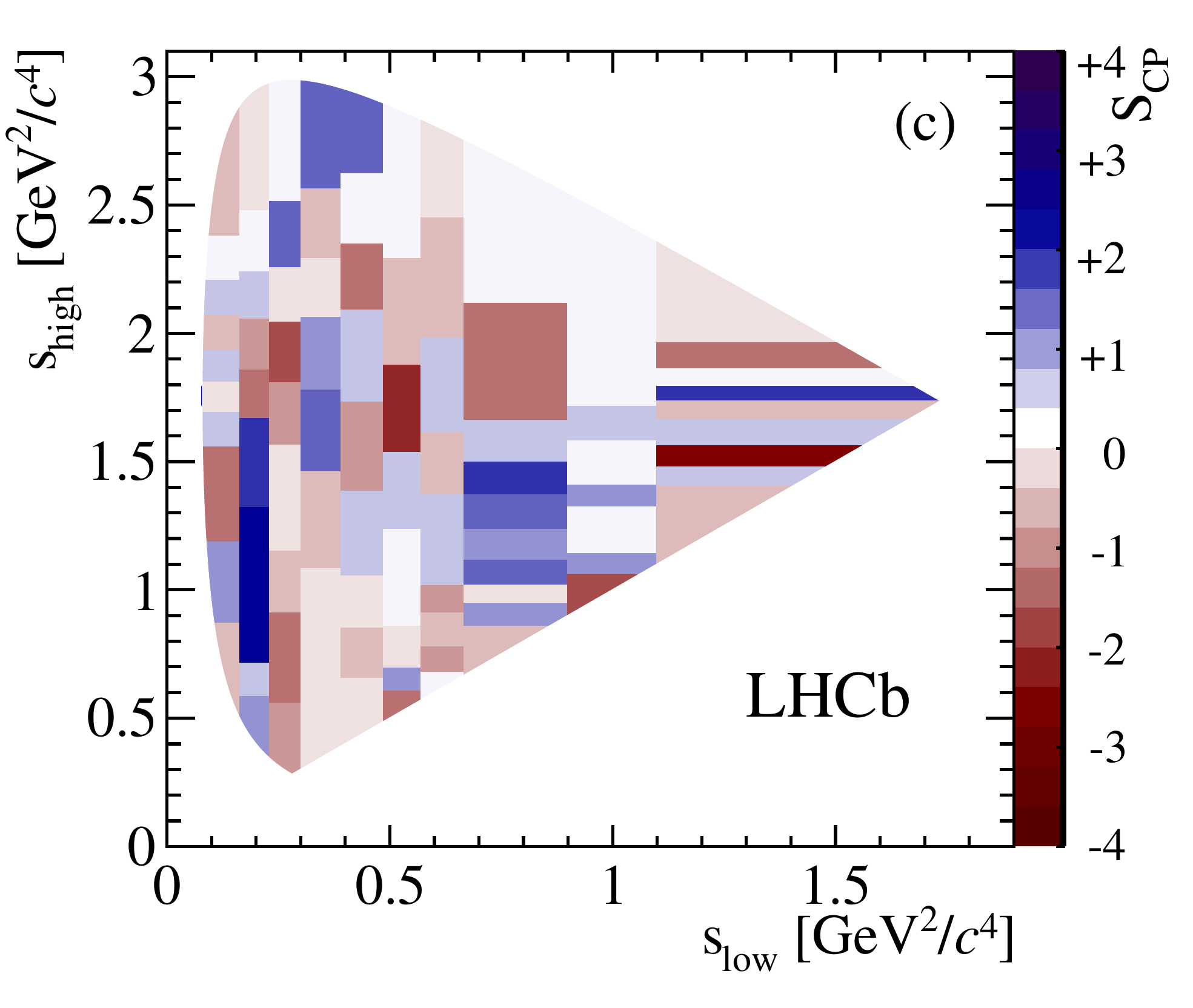}
\includegraphics[width=6.5cm]{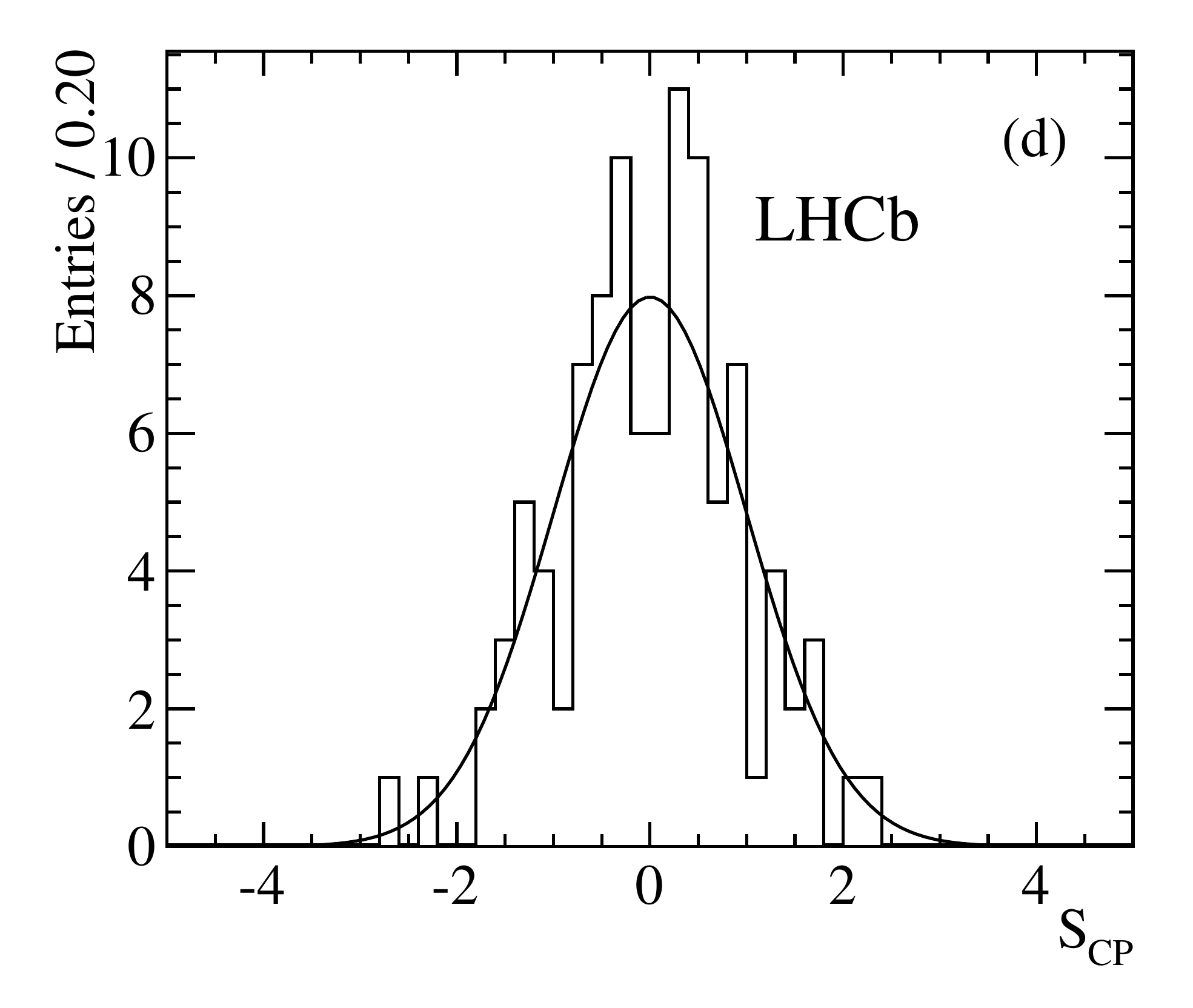}
\caption{The distribution of $S^i_{CP}$ across the $D^+ \to \pi^-\pi^+\pi^+$ 
Dalitz plot and the corresponding one-dimensional distributions.}
\label{fig:scp}
\end{figure}

Different bin divisions are tested. An adaptive scheme defines bins with 
equal population. As a cross-check, bins with equal size are also used.
In each scheme tests are performed using different number of bins. 
The distribution of $S^i_{CP}$ is shown in Fig. \ref{fig:scp}, together with
the distribution across the $D^+ \to \pi^-\pi^+\pi^+$, for two examples of
adaptive binning. No indication of \cpv\ was found.

\subsection{Summary}

Global fits to the existing measurements  were
performed HFAG collaboration~\cite{hfag}, 
extracting the mixing and \cp-violating parameters. The most general 
fit allows for all types of \cp\ violation. In the SM, however, direct 
\cp\ violation 
in DCS decays is not possible.  When this condition is imposed, from
the four mixing parameters --- $x$, $y$, $|q/p|$ and $\phi$ --- only three become
independent. A relation between the four parameters was obtained independently
by Chiuchini {\it et al.}~\cite{chiuc} and Kagan and Sokoloff~\cite{sokoloff}
(a slightly different formula was also obtained by Grossman {\it et al.}~\cite{gros09}), 
\begin{equation}
\tan \phi=\frac{x}{y}\left [\frac{1-|q/p|^2}{1+|q/p|^2}\right ]
\end{equation}
 
HFAG fit uses the formula to compute the averages. The impact of imposing this 
constraint is amazing, setting stringent bounds on $|q/p|$ and $\phi$. The
uncertainty in the value of $|q/p|$ allowing only for SM \cpv\ becomes 1.4\%,
and only a tenth of a degree for $\phi$. The result of the HFAG global 
fits are summarized in Table \ref{tab:averages}.

HFAG also fits for the underlying theory parameters,
\[
x_{12}\equiv \frac{2|M_{12}|}{\Gamma_D}, \hskip .4cm 
y_{12}\equiv \frac{|\Gamma_{12}|}{\Gamma_D}, \hskip .4cm 
\phi_{12}\equiv \arg\left(\frac{M_{12}}{\Gamma_{12}}\right):
\]
\begin{eqnarray}
x_{12} &=& (0.43\pm 0.14)\%,\nonumber\\
y_{12} &=& (0.60\pm0.07)\%,\nonumber\\
\phi_{12} &=& (0.9\pm1.6)^{\circ}.\nonumber 
\end{eqnarray}

\begin{figure}
\centering
\includegraphics[width=0.4\textwidth]{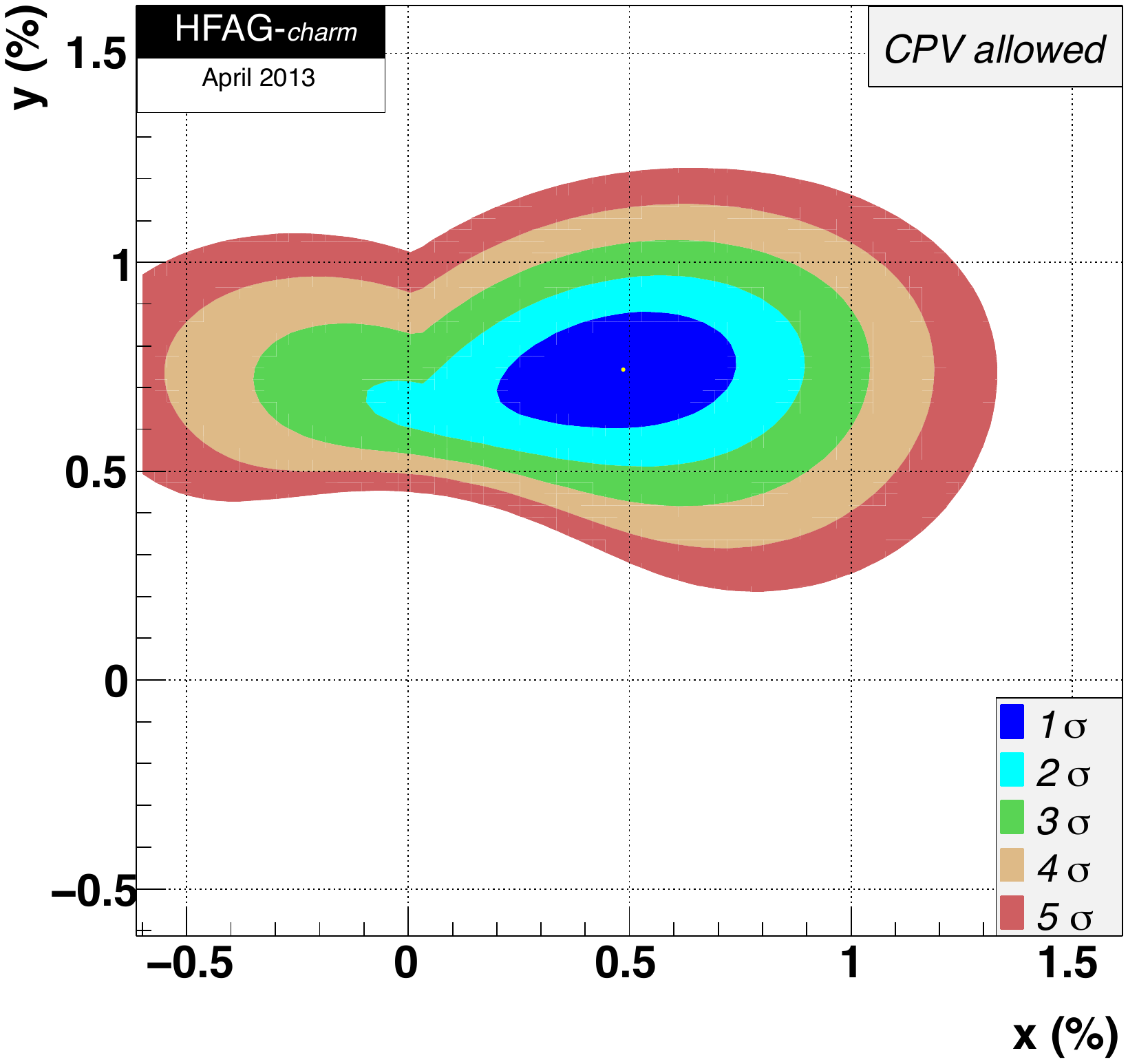}
\includegraphics[width=0.4\textwidth]{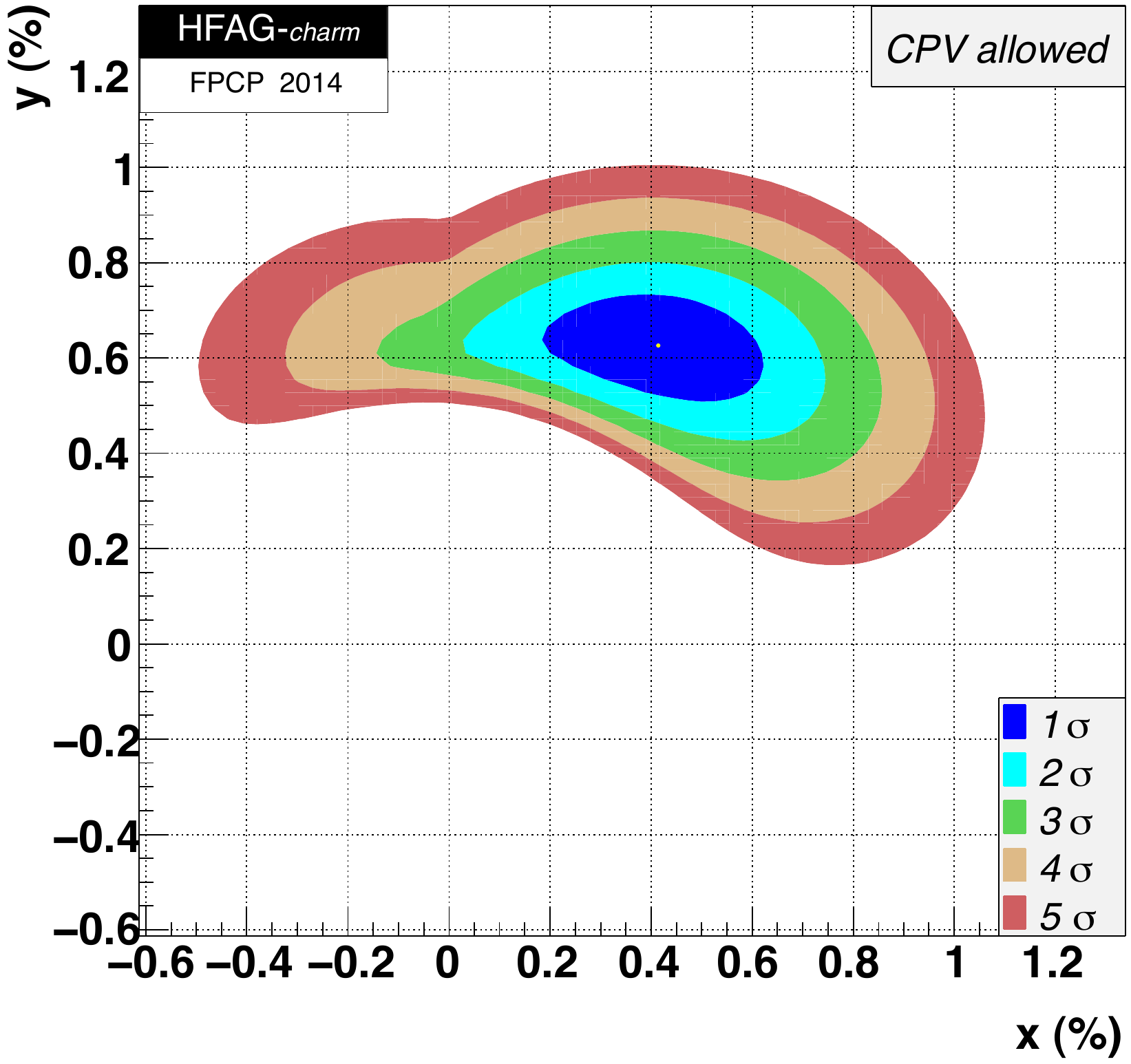}
\caption{Contour plots in the $x,y$ plane from a global fit to 41 measurements,
allowing for \cpv. On the left we see the confidence intervals from the fit
to the data available at April 2013. The plot on the right includes new measurements
since April 2013. Reproduced from HFAG~\cite{hfag}.}
\label{fig:hfagxy}
\end{figure}

\begin{figure}
\centerline{
\includegraphics[width=0.4\textwidth]{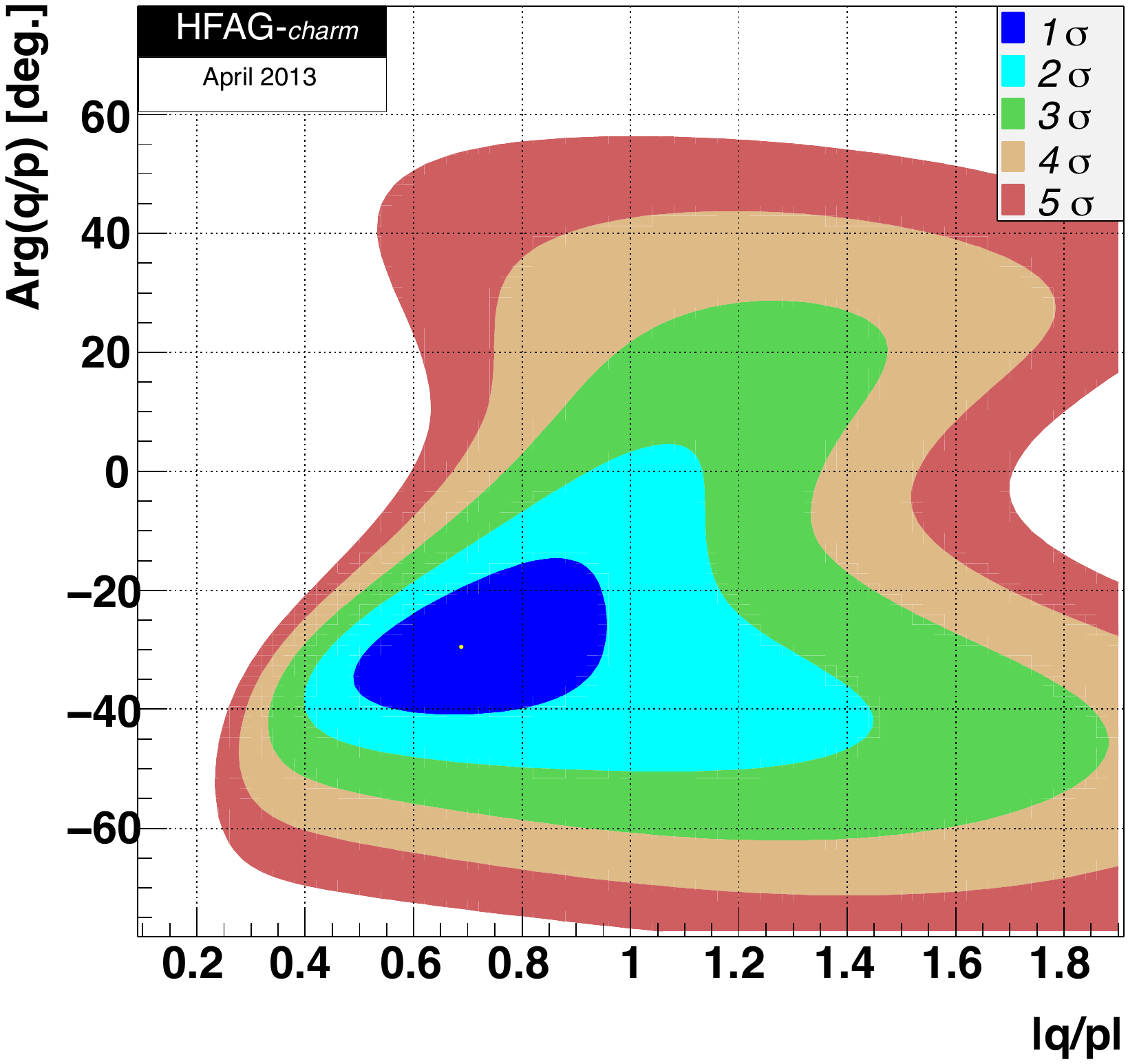}
\includegraphics[width=0.4\textwidth]{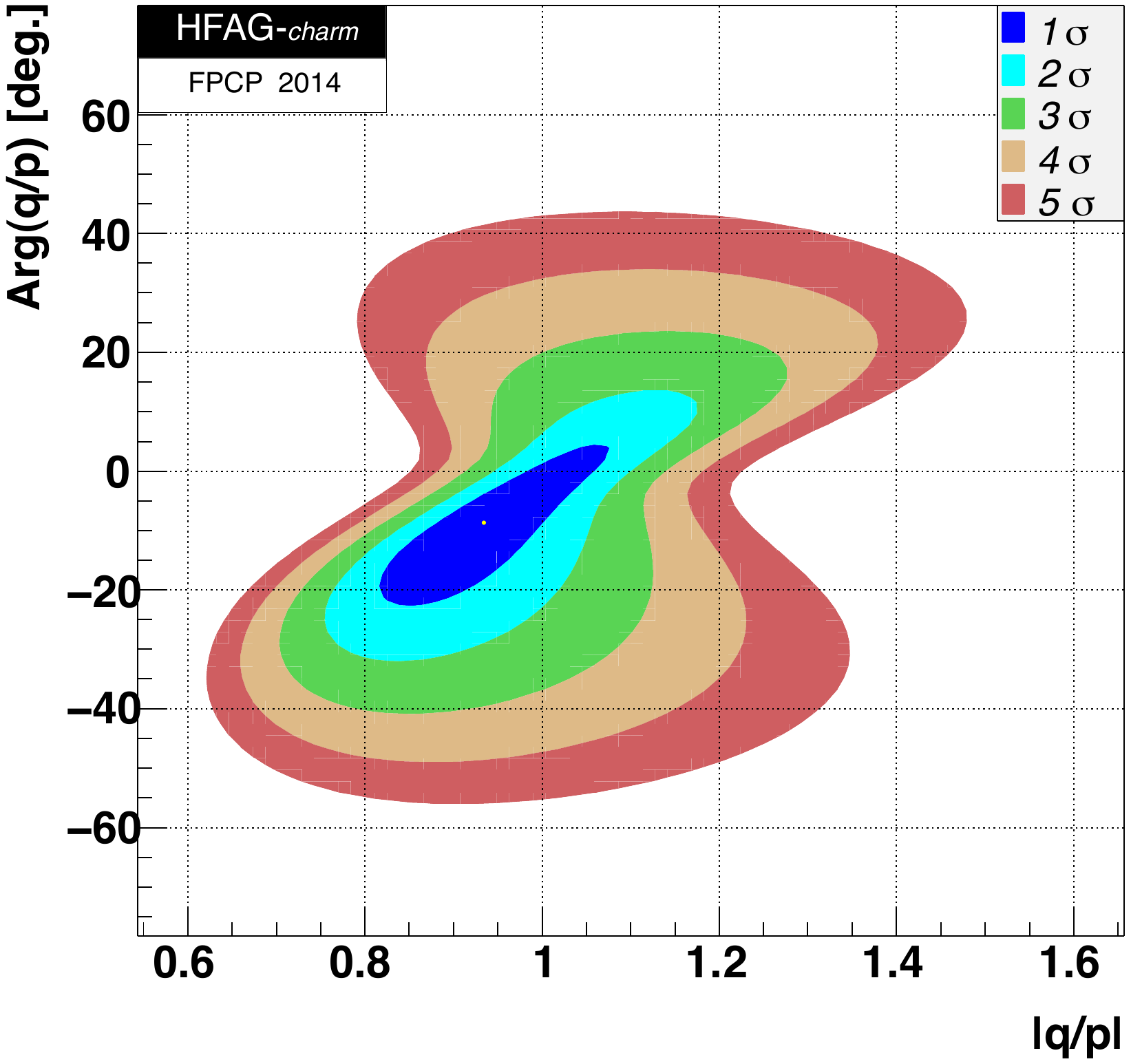}}
\caption{Contour plot of $|q/p|$ and its phase. On the left we see the confidence 
intervals from the fit to the data available at April 2013. The plot on the right 
includes new measurements since April 2013. 
Reproduced from HFAG~\protect\cite{hfag}.}
\label{fig:hfagqp}
\end{figure}

Contour plots with confidence level intervals from the \cpv-allowed fit 
are shown in Fig.~\ref{fig:hfagxy}, in the ($x,y$) plane, and in 
Fig.~\ref{fig:hfagqp}, in the $\phi,|q/p|$ plane. On the left panels 
contour plots from April 2013 are shown, whereas the right panel has the 
contours from May 2014. Note the change in the scale in both axes. 

From Fig.~\ref{fig:hfagxy} one sees that $y$ is better constrained than $x$. 
The sign of $y$ is well established. It is unlikely that $x$ is a negative 
quantity, but one should stress that its central value differs from zero by 
less than three standard deviations. A more precise value of $x$ will be
achieved when new measurements from LHCb become available.

An impressive improvement is observed in the confidence intervals of
the \cpv\ quantities $\phi$ and $|q/p|$, shown in Fig.~\ref{fig:hfagqp}, which
is mostly due to the LHCb results of the $D^0 \to K^{\mp}\pi^{\pm}$ decay.

Important measurements with the full data set from Run I are under 
way in LHCb: $y_{CP}$, $\Delta A_{CP}$ with pion tagged $D^0$, $x,y$ \ from
$D^0\to K_S^0\pi^-\pi^+$ time-dependent Dalitz plot analysis, $A_{\Gamma}$.
With data from Run II, starting in 2015, the study of mixing will enter a new
era of precision measurements. A significant improvement on the sensitivity 
of \cpv\ searches is also expected. However, for a correct interpretation of
the experimental results, advances in theory are necessary. In particular, 
the ability of computing the hadronic matrix elements is a crucial step,
allowing for a precise estimation of the SM contribution to both mixing and
\cp violation.

\begin{table}
\centering
\caption{Mixing and \cpv\ parameters obtained from a global fit of the existing 
measurements allowing for \cpv\ (from HFAG\cite{hfag}).}
{\begin{tabular}{ccc}
\\\hline
parameter       &    \cpv\ allowed             & no direct \cpv\ in DCS       \\
\hline
 $x$(\%)        & $0.41^{+0.14}_{-0.15}$      & $0.43^{+0.14}_{-0.15}$      \\
 $y$(\%)        & $0.63^{+0.07}_{-0.08}$      & $0.60^{+0.07}_{-0.08}$      \\  
$R_D$(\%)       & $0.3489^{+0.0038}_{-0.0037}$& $0.3485^{+0.0038}_{-0.0037}$\\
$A_D$(\%)       & $-0.71^{+0.92}_{-0.95}$     &  -                          \\       
$|q/p|$         & $0.93^{+0.08}_{-0.09}$      & $1.007^{+0.014}_{-0.015}$   \\
$\phi(^{\circ})$& $-8.7^{9.1}_{-8.8}$         & $-0.03^{0.10}_{-0.11}$      \\
\hline
\end{tabular}\protect\label{tab:averages}}
\end{table}

\newpage

\Acknowledgements
I am grateful to the Brazilian Conselho Nacional de Desenvolvimento 
Cient\'{\i}fico e Tecnol\'ogico for partially supporting this work.

\end{document}

%% file: econfmacros.tex
\def\beq{\begin{equation}}
\def\eeq#1{\label{#1}\end{equation}}
\def\eeqn{\end{equation}}

\def\beqa{\begin{eqnarray}}
\def\eeqa#1{\label{#1}\end{eqnarray}}
\def\eeqan{\end{eqnarray}}

\let\bar=\overbar

\def\Dslash{\not{\hbox{\kern-4pt $D$}}}
\def\dslash{\not{\hbox{\kern-2pt $\del$}}}

\def\cp{{\em CP }}

\def\msb{{\bar{\ssstyle M \kern -1pt S}}}

\def\ddbar{$D^0\!-\!\overline{D}^0$\ }
\def\cpv{{\em CP} violation}

